\documentclass[12pt,twoside]{article} 
\usepackage[latin1]{inputenc}
\usepackage{graphicx}
\usepackage{color}

\usepackage{enumitem}
\usepackage{latexsym}
\usepackage{amssymb}
\usepackage{epsfig}
\usepackage{hyperref}
\usepackage{xcolor}
\usepackage{amsmath}
\usepackage{natbib}
\usepackage{float}
\usepackage{dsfont}
\usepackage{multirow}
\usepackage{lscape}

\allowdisplaybreaks 

\newcommand{\Var}{\mbox{Var}}
\def\eps{\varepsilon}

\def \Cov{\mbox{Cov}}

\def\e{\varepsilon}

\def\beq{\begin{eqnarray*}}
\def\eeq{\end{eqnarray*}}

\def\t{\theta}

\newtheorem{theo}{Theorem}[section]
\newtheorem{lemma}[theo]{Lemma}
\newtheorem{cor}[theo]{Corollary}
\newtheorem{rem}[theo]{Remark}

\begin{document}

\begin{titlepage}

\title{\bf Specification tests in semiparametric transformation models -- a multiplier bootstrap approach}

\author{{\sc Nick Kloodt} and {\sc Natalie Neumeyer\footnote{Corresponding author; e-mail: neumeyer@math.uni-hamburg.de; financial support by the DFG  (Research Unit FOR 1735 {\it Structural Inference in
Statistics: Adaptation and Effciency}) is gratefully acknowledged.}}\\ Department of Mathematics, University of Hamburg}

\maketitle

\begin{abstract}
We consider semiparametric transformation models, where after pre-estimation of a parametric transformation of the response the data are modeled by means of nonparametric regression. 
We suggest subsequent procedures for testing lack-of-fit of the regression function and for significance of covariables, which -- in contrast to procedures from the literature -- are asymptotically not influenced by the pre-estimation of the transformation. The test statistics are asymptotically pivotal and have the same asymptotic distribution as in regression models without transformation. We show validity of a multiplier bootstrap procedure which is easier to implement and much  less computationally demanding than bootstrap procedures based on the transformation model. In a simulation study we demonstrate the superior performance of the procedure in comparison with the competitors from the literature. 
\end{abstract}

\noindent{\bf Key words:} Box-Cox transformation, lack-of-fit test, multiplier bootstrap, nonparametric regression, significance of covariables, U-statistics,  Yeo-Johnson transformation

\noindent{\bf AMS 2010 Classification:}  62G08, 62G09, 62G10

\end{titlepage}

\section{Introduction} 

Assume we have observed independent data $(X_i,Y_i)$, $i=1,\dots,n$, and after a transformation of the response $Y_i$ a regression model shall be fitted. 
The aim of the data transformation typically is to obtain a simpler model, e.g.\ a homoscedastic instead of a heteroscedastic model. 
If the transformation is chosen from a parametric class $\{\Lambda_\t\mid \t\in\Theta\}$ (e.g.\ Box-Cox power transformations, see Box and Cox, 1964, or their modification suggested by Yeo and Johnson, 2000) one typically assumes the existence of a unique `true parameter' $\t_0\in\Theta$ such that the simpler model holds for the transformed data $(X_i,\Lambda_{\t_0}(Y_i))$, $i=1,\dots,n$. We will assume a homoscedastic model 
\begin{eqnarray*}
\Lambda_{\t_0}(Y_i) &=& m(X_i)+\eps_i,\quad i=1,\dots,n,
\end{eqnarray*}
where $m$ denotes the regression function and $\eps_i$ some unobservable centered error, independent of the covariates $X_i$. 
Data-dependent choices of the transformation parameter as considered by  Linton, Sperlich and Van Keilegom (2008), among others, however, leave subsequent inference to be based on 
$(X_i,\Lambda_{\hat\t}(Y_i))$, $i=1,\dots,n$, 
where $\hat\t$ depends on the whole sample $(X_1,Y_1),\dots,(X_n,Y_n)$. 
It would be desirable to be able to apply standard procedures to the transformed data. However, the random transformation may influence the performance of inference procedures severely. 
For the famous Box-Cox transformations this phenomenon has been a major discussion topic (see, e.g.\ Bickel and Doksum, 1981, or Hinkley and Runger, 1984), but is often ignored in practice. Note that estimating the transformation is a problem related to model selection. The influence of the random selection procedure on post-model-selection inference is a topic of current high interest, see, e.g.\ Berk et al.\ (2013), Efron (2014), Lee et al.\ (2016) or Charkhi and Claeskens (2018).
Naturally inferential procedures that are not influenced by the randomness of the model selection (here estimation of the transformation parameter) have the advantage of ready applicability. 

For the model at hand we will consider two typical testing problems in nonparametric regression models, namely testing for a parametric class of the regression function (lack-of-fit) and testing for significance of covariables. We will present test statistics that are asymptotically not influenced by the randomness of the data transformation. 

Concerning the first testing problem, recently Colling and Van Keilegom (2016, 2017) suggested lack-of-fit tests for the regression function $m$ in a transformation model. The tests are based on ideas from Van Keilegom, Gonzßlez-Manteiga and Sßnchez Sellero (2008), on the one hand, and from Bierens (1982), Stute (1997) and Escanciano (2006), on the other hand (in models without transformation). Colling and Van Keilegom (2016, 2017) derive the asymptotic distribution of the test statistics and show that the estimation of the transformation parameter alters the limit distribution in both cases. Even when bootstrap is conducted to apply the tests, the bootstrap versions of the original tests (without transformation) cannot be applied to the transformed data, but the bootstrap procedures have to be adapted for the transformation model as well. In particular, for each bootstrap replication a new transformation estimation has to be performed. This can be computationally quite demanding as the estimation is based on nonlinear optimization. 

Concerning the second testing problem, Allison, Hu\v{s}kova and Meintanis (2018) consider testing for significance of covariables in semiparametric transformation models based on ideas from  Bierens (1982) and Hlßvka, Hu\v{s}kovß,  Kirch and  Meintanis (2017). They derive the asymptotic distribution of the test statistic, which again is influenced by the transformation estimation. The bootstrap procedure is thus adapted to the unknown transformation. 

The changes in the asymptotic distributions due to the transformation estimation and the necessary modification of standard procedures seems to be rather inconvenient for applications. They are contrary to the expectation that with the transformation one obtains a simple model, for which standard inference procedures can be applied. 
On account of this we will suggest testing procedures that are asymptotically not influenced by the random transformation. Therefore, after the data transformation standard procedures or their standard bootstrap versions can be applied. For testing for a parametric class of the regression functions we will generalize Hõrdle and Mammen's (1993) test (see also Alcalß, Crist¾bal and Gonzßlez-Manteiga, 1999) as well as Zheng's (1996) test. For testing significance of covariables we will generalize Lavergne, Maistre and  Patilea's (2015) procedure. 
We moreover suggest multiplier bootstrap versions of the tests and show validity of this approach. The multiplier bootstrap is much easier to implement as well as faster than the transformation-model-based bootstrap used in the literature. Van der Vaart and Wellner (1996) describe multiplier bootstrap methods; see also Zhu et al.\ (2001), B³cher and Dette (2013), Spokoiny and Zhilova (2015), among others. 

In section 2 we will define the semiparametric transformation model and briefly discuss the estimation of the transformation parameter. In section 3 we suggest two lack-of-fit tests for the regression function after transformation, while in section 4 we consider testing for significance of covariables. In both settings we prove asymptotic normality of our test statistics under the null hypothesis and local alternatives, suggest multiplier bootstrap versions of the tests and show asymptotic validity of this approach. 
In section 5 we compare the suggested tests with those from the literature in a simulation study. 
Appendix A contains the assumptions and Appendix B the proofs.

\section{Estimation of the transformation}
\def\theequation{2.\arabic{equation}}
\setcounter{equation}{0}

Throughout we assume we have independent realisations $(X_1,Y_1),...,(X_n,Y_n)$ of the model
\begin{equation}\label{model}
\Lambda_{\theta_0}(Y)=m(X)+\varepsilon,
\end{equation} 
where $Y$ and $\varepsilon$ are $\mathbb{R}$-valued and $X$ is $\mathbb{R}^d$-valued random variable with density $f_X$. Moreover,  $\varepsilon$ and $X$ are independent and $E[\varepsilon]=0$, $\Var(\varepsilon)=\sigma^2\in (0,\infty)$. The transformation belongs to a parametric class $\{\Lambda_\theta\mid\theta\in\Theta\}$ of strictly increasing functions, and the true parameter $\theta_0$ is unknown. 
The regression function $m$ is estimated nonparametrically. 

There are several possibilities to estimate the transformation parameter. For our approaches the only property the estimator $\hat\theta$ has to fulfill is root-$n$-consistency, i.e.\
\begin{equation}\label{hat-theta}
\hat\theta=\theta_0+O_P(n^{-1/2}).
\end{equation} 
This is fulfilled for the profile likelihood and the minimum distance estimator considered in Linton et al.\ (2008). See Colling and Van Keilegom (2016) for a detailed description of the profile likelihood estimator that we use in our simulations, and for regularity assumptions to obtain (\ref{hat-theta}). 
Recently, Colling and  Van Keilegom (2018) suggested an alternative estimator for $\theta$ that also fulfills (\ref{hat-theta}). 

For transformation parameter estimation in other semiparametric models, see, e.g.\ \linebreak Horowitz (1996) or Linton, Chen, Wang and Hõrdle (1997). Some parametric classes for transformations are considered by Box and Cox (1964), Zellner and
Revankar (1969), Bickel and Doksum (1981) and Yeo and Johnson (2000), among others.

\section{Lack-of-fit testing}\label{section-gof}
\def\theequation{3.\arabic{equation}}
\setcounter{equation}{0}

\subsection{Hypotheses, test statistics and asymptotic distribution}

For model (\ref{model}) we consider tests for the hypothesis
$$H_0:\quad m\in \{m(\cdot,\beta):\beta\in B\}$$
for some $q$-dimensional compact parameter space  $B$. The function $m(\cdot,\beta)$ is known apart from the true regression parameter $\beta_0$ with $m(\cdot)=m(\cdot,\beta_0)$. 
Testing lack-of-fit (in models without transformation) is a classical topic in statistics. A very thorough review on related literature is given in Gonzßlez-Manteiga and Crujeiras (2013). The most commonly used approaches (that also have been very influential in terms of development of related statistics in different contexts) are arguably those by Hõrdle and Mammen (1993) (based on $L^2$ distance), Zheng (1996) (based on $U$-statistics) and Stute (1997) (based on a marked empirical process). We take the two first approaches in a model with unknown transformation, while the latter one was considered by Colling and Van Keilegom (2017).

To this end we consider least squares estimators for the regression parameter based on the transformed data $(X,\Lambda_\theta(Y))$ for each $\theta\in\Theta$,
$$\hat{\beta}_{\theta}=\underset{\beta\in\,B}{\operatorname{argmin}}\sum_{i=1}^n(\Lambda_{\theta}(Y_i)-m(X_i,\beta))^2.$$
We further define  local alternatives
\begin{equation}\label{loc-alt}m(x)=m(x,\bar{\beta})+c_n\Delta_n(x)
\end{equation}
with parameter $\bar{\beta}=\arg\min_{\beta\in B}\int(m(x)-m(x,\beta))^2f(x)\,dx$, rate $c_n\sim n^{-\frac{1}{2}}h^{-\frac{d}{4}}$ and $\Delta_n(x)$ uniformly bounded in $n$ and $x$. Note that $\bar\beta$ may depend on $n$. These local alternatives contain as special case the null hypothesis with $\Delta_n\equiv 0$ and $\bar{\beta}=\beta_0$.
Now we define Hõrdle and Mammen's (1993) test statistic 
$$T_n(\theta)=nh^{\frac{d}{2}}\int\bigg(\frac{1}{n}\sum_{i=1}^nK_h(x-X_i)(\Lambda_{{\theta}}(Y_i)-m(X_i,{\hat{\beta}_{{\theta}}}))\bigg)^2\,dx$$
and 
Zheng's (1996) test statistic 
$$V_n(\theta)=\frac{1}{n(n-1)}\sum_{i=1}^n\sum_{\stackrel{\scriptstyle j=1}{j\neq i}}^nK_h(X_i-X_j)(\Lambda_{{\theta}}(Y_i)-m(X_i,{\hat{\beta}_{{\theta}}}))(\Lambda_{{\theta}}(Y_j)-m(X_j,\hat{\beta}_{{\theta}}))$$
both applied to the transformed data $(X,\Lambda_\theta(Y))$. 
Here $K_h(\cdot)=K(\cdot/h)/h^d$, where $K:\mathbb{R}^d\to\mathbb{R}$ denotes a kernel function and $h=h_n$ a sequence of bandwidths fulfilling assumptions \ref{A3} and \ref{A4} in Appendix A. 
Note that in contrast to Hõrdle and Mammen (1993) we avoid the kernel density estimator in the denominator (see also Alcalß, Crist¾bal and Gonzßlez-Manteiga, 1999). 
As test statistics we consider $T_n(\hat\theta)$ and $V_n(\hat\theta)$ for some estimator $\hat\theta$ that fulfills (\ref{hat-theta}), whereas the asymptotic distributions of $T_n(\theta_0)$ and  $V_n(\theta_0)$  are given by Hõrdle and Mammen (1993) and Zheng (1996), respectively (in a model without transformation).

\begin{theo}\label{theo1} Under the assumptions \ref{A1}--\ref{A9} in Appendix A, we have under the local alternatives (\ref{loc-alt}), 
$T_n(\hat\theta)-T_n(\theta_0)=o_P(1)$ and $V_n(\hat\theta)-V_n(\theta_0)=o_P((nh^{d/2})^{-1})$. 
\end{theo}

The proof is given in Appendix \ref{prooftheo1}. Theorem \ref{theo1} shows that the asymptotic distribution is not influenced by the estimation of the transformation. The reason is essentially the faster convergence of the transformation parameter estimator compared to the convergence rate of the test statistics. 
From Hõrdle and Mammen (1993) and Zheng (1996) now directly follows the next result. 

\begin{cor}\label{cor-gof}
Under the assumptions of Theorem \ref{theo1} we have under the local alternatives (\ref{loc-alt}) for $n\to\infty$, 
\begin{eqnarray*}
\frac{T_n(\hat\theta)-b_h-\mu_n}{\sqrt{V}}&\overset{\mathcal{D}}{\longrightarrow}&\mathcal{N}(0,1)\\
\frac{nh^{\frac{d}{2}}V_n(\hat\theta)-\mu_n}{\sqrt{\Sigma} } &\overset{\mathcal{D}}{\longrightarrow}&\mathcal{N}(0,1)
\end{eqnarray*}
with $b_h=h^{-\frac{d}{2}}\sigma^2\int K^2(u)\,du$, $\mu_n=E[\Delta_n(X_1)^2f_X(X_1)]$, $V=2\sigma^4\int f_X(x)^2\,dx\int (K*K)^2(x)\,dx$, where $(K*K)(x)=\int K(x-u)K(u)\,du$,  and $\Sigma=2\sigma^4\int f_X(x)^2\,dx\int K^2(u)\,du$. 
\end{cor}

Asymptotic level $\alpha$-tests for the null hypothesis $H_0$ can be constructed from Corollary \ref{cor-gof}. 
To this end, note that with the methods used in the proof of Theorem \ref{theo1}, it is easy to show that consistent estimators  (under the null) for $\sigma^2$ and $\Sigma$ are given by 
$\hat\sigma^2=n^{-1}\sum_{i=1}^n(\Lambda_{\hat{\theta}}(Y_i)-m(X_i,\hat\beta_{\hat\theta}))^2$
and $\hat\Sigma(\hat\theta)$, respectively, with 
$$\hat{\Sigma}(\theta)=\frac{2}{n(n-1)h^d}\sum_{i=1}^n\sum_{\stackrel{\scriptstyle j=1}{j\neq i}}^nK^2\Big(\frac{X_i-X_j}{h}\Big)(\Lambda_{{\theta}}(Y_i)-m(X_i,{\hat{\beta}_{{\theta}}}))^2(\Lambda_{{\theta}}(Y_j)-m(X_j,\hat{\beta}_{{\theta}}))^2.$$
Further $\hat{s}=n^{-2}\sum_{i=1}^n\sum_{j=1}^nK_h(X_j-X_i)$ consistently estimates  $\int f_X(x)^2\, dx$.
Note that Zheng (1996) uses the estimator $\hat\Sigma(\theta_0)$ in a heteroscedastic model. In our model, instead of $\hat\Sigma(\hat\theta)$, the estimator $\tilde\Sigma=2\hat\sigma^4\hat s\int K^2(y)\,dy$ can be applied as well. 
Consistency of the asymptotic tests  follows because $T_n(\hat\theta)-b_h$ as well as $nh^{\frac{d}{2}}V_n(\hat\theta)$ diverge to infinity under fixed alternatives. 
We demonstrate the finite sample behavior of the asymptotic tests as well as of several bootstrap versions in section \ref{simus}. 

Corollary \ref{cor-gof} further shows that both tests can detect local alternatives of rate $(nh^{d/2})^{-1/2}$.
 The approaches considered by Colling and Van Keilegom (2016, 2017) can detect faster local alternatives of $n^{-1/2}$-rate, but the asymptotic distributions depend in a complicated way on the estimation of the transformation. Further, the estimation of the transformation needs also to be taken into account in the bootstrap procedure. 

\begin{rem}
(i) 
For goodness-of-fit tests sometimes it is argued that empirical process based tests (as introduced by Stute, 1997) should be preferred over smoothing based tests (like those suggested by Hõrdle and Mammen, 1993, and Zheng, 1996), because the latter introduce the choice of a smoothing parameter (to estimate the regression function nonparametrically). Note that this is not a relevant argument in the context of transformation models, because the choice of the smoothing parameter for estimating $m$ is already needed in order to estimate the transformation parameter and thus also necessary for the empirical process based tests. 

(ii)
Other test statistics could be considered as well, e.g.\ the empirical characteristic function approach by Hu\v{s}kovß  and Meintanis (2009). We conjecture that with this approach the asymptotic distribution will depend on the transformation parameter estimation. 
\end{rem}

\subsection{Bootstrap versions}\label{bootstrapversions}

Hõrdle and Mammen (1993) already noticed that due to the slow convergence rate of the negligible terms the asymptotic distribution may be inappropriate for obtaining critical values. Hence, they suggested a wild bootstrap procedure that is based on the ideas of Wu (1986).
We will call this approach `standard wild bootstrap' (swb) in what follows. To describe it in our context denote the transformed data as $(X_i,Z_i)$, $i=1,\dots,n$, with $Z_i=\Lambda_{\hat{\theta}}(Y_i)$, and define residuals $\hat{\varepsilon}_i=Z_i-\hat{m}(X_i)$ as nonparametric estimates of the errors, 
where $\hat m $ denotes the Nadaraya-Watson estimator for ${m}$ based on the sample $(X_i,Z_i)$, $i=1,\dots,n$ (see Nadaraya, 1964, or Watson, 1964). Further, generate  independent random variables $U_1,\dots,U_n$, independent from the sample, with expectation zero, unit variance and unit third moment. 
Now use the new sample $(X_i,Z_i^*=m(X_i,\hat{\beta}_{\hat{\theta}})+\hat\varepsilon_iU_i)$, $i=1,\dots,n$,  as bootstrap observations.  
The bootstrap versions of the test statistics are now defined as 
\begin{eqnarray*}
T_{n}^{\text{swb}*}&=&nh^{\frac{d}{2}}\int\bigg(\frac{1}{n}\sum_{i=1}^nK_h(x-X_i)(Z_i^*-m(X_i,{\hat{\beta}^*}))\bigg)^2\,dx\\
V_n^{\text{swb}*}&=&\frac{1}{n(n-1)}\sum_{i=1}^n\sum_{\stackrel{\scriptstyle j=1}{j\neq i}}^nK_h(X_i-X_j)(Z_i^*-m(X_i,\hat{\beta}^*))(Z_j^*-m(X_j,\hat{\beta}^*)),
\end{eqnarray*}
where $\hat\beta^*$ is evaluated from $(X_i,Z_i^*)$, $i=1,\dots,n$. 
Note that the bootstrap, in contrast to the one presented by Colling and Van Keilegom (2017), 
does not take account of the estimation of the transformation parameter. It nevertheless leads to an asymptotically valid procedure due to the asymptotic negligibility of the transformation parameter estimation. However, for small and moderate sample sizes we will see in the simulation section that the level is often overestimated. The reason is that due to the estimation error of $\hat\theta$ the data $(X_i,Z_i)$ do typically not exactly fulfill the null model, whereas the bootstrap data $(X_i,Z_i^*)$ do. 

Colling and Van Keilegom (2017) take into account the transformation estimation in their bootstrap procedure (because in their context the standard wild bootstrap as above does not lead to asymptotically valid procedures). 
We will call this approach `transformation wild bootstrap' (twb) in what follows. 
 To this end define $\varepsilon_i^*:=\zeta_i^*+a_n\xi_i^*$, where the $\xi_i^*$ are standard normally distributed random variables independent of $(X_i,Y_i)$, $i=1,\dots,n$,  the $\zeta_i^*$ are drawn with replacement from the nonparametrically estimated residuals, and $a_n\to 0$.  Then the bootstrap sample $(X_i^*,Y_i^*)$ is obtained by $X_i^*=X_i$ and $Y_i^*=\Lambda_{{\hat\theta}}^{-1}(m(X_{i}^*,\hat{\beta})+\varepsilon_i^*)$. The bootstrap versions of our test statistics are  defined as 
\begin{eqnarray*}
T_n^{\text{twb}*}&=&nh^{\frac{d}{2}}\int\bigg(\frac{1}{n}\sum_{i=1}^nK_h(x-X_i)(\Lambda_{{\hat\theta^*}}(Y_i^*)-m(X_i,{\hat{\beta}_{{\hat\theta^*}}}))\bigg)^2\,dx\\
V_n^{\text{twb}*}&=&\frac{1}{n(n-1)}\sum_{i=1}^n\sum_{\stackrel{\scriptstyle j=1}{j\neq i}}^nK_h(X_i-X_j)(\Lambda_{{\hat\theta^*}}(Y_i^*)-m(X_i,{\hat{\beta}_{{\hat\theta^*}}}))(\Lambda_{{\hat\theta^*}}(Y_j^*)-m(X_j,\hat{\beta}_{{\hat\theta^*}})),
\end{eqnarray*}
where $\hat\theta^*$ is the transformation parameter estimator built from the bootstrap sample.
This approach is much more computationally demanding than the standard wild bootstrap, but does lead to better approximation of the level. 

In what follows we will introduce an alternative approach: the multiplier bootstrap, which is easy to implement, not computationally demanding and leads to good level approximations and high power. To this end note that with parametric residuals $\hat e_i= \Lambda_{\hat{\theta}}(Y_i)-m(X_i,{\hat{\beta}_{\hat{\theta}}})$, $i=1,\dots,n$, we can write 
$$T_n(\hat{\t})=\frac{h^{\frac{d}{2}}}{n}\sum_{i=1}^n\sum_{j=1}^n(K\ast K)_h(X_i-X_j)\hat e_i\hat e_j,$$
where $(K\ast K)_h(z)=\int K_h(x)K_h(z-x)\,dx$ denotes the convolution of $K_h$, and
$$V_n(\hat{\t})=\frac{1}{n(n-1)}\sum_{i=1}^n\sum_{\stackrel{\scriptstyle j=1}{j\neq i}}^nK_h(X_i-X_j)\hat e_i\hat e_j.$$
Now let $\xi_1,...,\xi_n$ be iid random variables independent of $(Y_i,X_i),i=1,...,n$, with expectation zero, variance one and existing fourth moments. Define the multiplier bootstrap test statistic as
\begin{eqnarray*}
T_n^{\text{mb}*} &=& \frac{h^{\frac{d}{2}}}{n}\sum_{i=1}^n\sum_{j=1}^n(K\ast K)_h(X_i-X_j)\hat e_i\hat e_j\xi_i\xi_j\\
V_n^{\text{mb}*}&=&\frac{1}{n(n-1)}\sum_{i=1}^n\sum_{\stackrel{\scriptstyle j=1}{j\neq i}}^nK_h(X_i-X_j)\hat e_i\hat e_j\xi_i\xi_j.
\end{eqnarray*}
(notation `mb' for multiplier bootstrap), and `centered versions' as 
 \begin{eqnarray*}
T_n^{\text{cmb}*} &=& \frac{h^{\frac{d}{2}}}{n}\sum_{i=1}^n\sum_{j=1}^n(K\ast K)_h(X_i-X_j)(\hat e_i\xi_i-\overline{\hat e\xi}_n)(\hat e_i\xi_j-\overline{\hat e\xi}_n)\\
V_n^{\text{cmb}*}&=&\frac{1}{n(n-1)}\sum_{i=1}^n\sum_{\stackrel{\scriptstyle j=1}{j\neq i}}^nK_h(X_i-X_j)(\hat e_i\xi_i-\overline{\hat e\xi}_n)(\hat e_i\xi_j-\overline{\hat e\xi}_n)
\end{eqnarray*}
(notation `cmb'), where we define ${\overline{\hat e\xi}}_n=n^{-1}\sum_{i=1}^n \hat e_j\xi_i$.

The next theorem establishes the asymptotic properties of the bootstrap test statistics. It is valid under the null hypothesis as well as under fixed alternatives. To this end let $\bar\beta=\arg\underset{\beta\in B}{\min}E[(m(X)-m(X,\beta))^2]$ and define $\Delta(x)=m(x)-m(x,\bar\beta)$, which vanishes under $H_0$.

\begin{theo}\label{theo-boot-gof}
Under the assumptions \ref{A1}--\ref{A9} in Appendix A (see Remark \ref{Delta-fixed}) for $n\to\infty$
$$\frac{T_n^{\text{mb}*}-b_h^*}{\sqrt{V^*}} \;\mbox{ and }\;\frac{nh^{\frac{d}{2}}V_n^{\text{mb}*}}{\sqrt{\Sigma^*} },$$
conditionally on $(X_1,Y_1),\dots,(X_n,Y_n)$, converge in distribution to standard normal distributions, in probability. The same holds with $T_n^{\text{mb}*}$ replaced by $T_n^{\text{cmb}*}$ and $V_n^{\text{mb}*}$ replaced by $V_n^{\text{cmb}*}$. 
Here,  $b_h^*=h^{-\frac{d}{2}}E[(\e_1+\Delta(X_1))^2]\int K(x)^2\,dx$ coincides with $b_h$ under $H_0$, $V^*=2\int(\sigma^2+\Delta(x)^2)^2f(x)^2\,dx\int(K\ast K)^2(u)\,du$ coincides with $V$ under $H_0$, and $\Sigma^*=2\int(\sigma^2+\Delta(x)^2)^2f_X(x)\,dx\int K^2(u)\,du$ coincides with $\Sigma$ under $H_0$. 
\end{theo}

The proof is given in Appendix \ref{beweis-boot}.
From Theorem \ref{theo-boot-gof} it follows that approximating the critical values from the multiplier bootstrap versions leads to consistent asymptotic level $\alpha$ tests.

\begin{rem}
The multiplier bootstrap cannot be applied to Colling and Van Keilegom's (2016, 2017) procedures in any obvious way. For instance, consider the modification of Stute's (1997) test statistic for the transformation model, which is based on the process $S_n(x)=n^{-1/2}\sum_{i=1}^n \hat e_iI\{X_i\leq x\}$. A multiplier bootstrap version could be defined as \linebreak $S_n^{\text{mb}*}(x)=n^{-1/2}\sum_{i=1}^n \hat e_iI\{X_i\leq x\}\xi_i$. Then the conditional covariances, given the sample,  \linebreak $\Cov^*(S_n^{\text{mb}*}(x), S_n^{\text{mb}*}(z))=n^{-1}\sum_{i=1}^n \hat e_i^2I\{X_i\leq x\wedge z\}$ converge under $H_0$ in probability to $\sigma^2 F_X(x\wedge z)$, which does not coincide with the asymptotic covariance between $S_n(x)$ and $S_n(z)$.
\end{rem}

\section{Testing significance of covariables}\label{testsignificance}
\def\theequation{4.\arabic{equation}}
\setcounter{equation}{0}

\subsection{Hypotheses, test statistics and asymptotic distribution}

Under model (\ref{model}) let $X=(W,V)$, where the entries of $X$ are in such order that the hypothesis of significance of $V$ is of interest, i.e.
$$H_0:  E[\Lambda_{\theta_0}(Y)|W,V]= E[\Lambda_{\theta_0}(Y)|W]. $$
Note that $m(X)=m(W,V)= E[\Lambda_{\theta_0}(Y)|W,V]$ and throughout we use the definition $r(W)=E[\Lambda_{\theta_0}(Y)|W]$. Further, we will denote the density of $W$ by $f_W$ and the conditional density of $W$, given $V=v$, by $f_{W|V}(\cdot|v)$. 

Literature overviews for testing for the simplifying hypothesis $H_0$ (in models without transformation) are given by Gonzßlez-Manteiga and Crujeiras (2013) and Lavergne et al.\ (2015). Lavergne et al.'s (2015) test is similar to those by Fan and Li (1996) and Lavergne and Vuong (2000) based on U-statistics, but does not involve smoothing with respect to $V$ and therefore converges at a faster rate, independent from the dimension of $V$. 

Assume that  $W$ is $p$- and $V$ $q$-dimensional. Like Lavergne et al.\ (2015) we consider the local alternative
	\begin{equation}\label{loc-alt-sig}
m(W,V)=r(W)+\delta_nd(W,V)
\end{equation}
for some fixed integrable function $d:\mathbb{R}^{p+q}\rightarrow\mathbb{R}$ with $E[d(W,V)|W]=0$, but concentrate on the rate $\delta_n\sim n^{-\frac{1}{2}}h^{-\frac{p}{4}}$.
The local alternative contains the null hypothesis for $d=0$. 

 We define  
	\begin{align*}
	{I}_n (\theta)&=\frac{h^{\frac{p}{2}}}{n^3}{\sum_{i,j,k,l}}^{\neq}\left(\Lambda_\theta(Y_i)-\Lambda_\theta(Y_k)\right)\left(\Lambda_\theta(Y_j)-\Lambda_\theta(Y_l)\right)L_g(W_i-W_k)L_g(W_j-W_l)
	\\[0,2cm]&\qquad\qquad\qquad\qquad  \times K_h(W_i-W_j)\psi(V_i-V_j).
	\end{align*}
As before, $K_h(\cdot)=K(\cdot/h)/h^p$ and $L_g(\cdot)=L(\cdot/g)/g^p$, where $K$ and $L$ denote bounded and symmetric kernel functions of $p$ variables with compact supports together with two bandwidths $g=g_n$ and $h=h_n$. Let $\psi$ be a bounded, symmetric function with almost everywhere positive Fourier transform. 
The notation $\sum^{\neq}$ stands for summation over pairwise distinct indices. 

Using a transformation parameter estimator $\hat\theta$ as in (\ref{hat-theta}),  we consider $ I_n(\hat\theta)$ as test statistic, which is asymptotically equivalent to $I_n(\theta_0)$ by the following theorem.

\begin{theo}\label{theo2} Under the assumptions \ref{B1}--\ref{B11} in Appendix A,  we have under the local alternatives (\ref{loc-alt-sig})
$ I_n(\hat\theta)- I_n(\theta_0)=o_P(1)$.
\end{theo}

The proof of Theorem \ref{theo2} is given in Appendix \ref{prooftheo2}, while the following Corollary \ref{cor-theo2} is a direct consequence of Lavergne et al.\ (2015). 

\begin{cor}\label{cor-theo2}
Under the assumptions of Theorem \ref{theo2} we have under the local alternatives (\ref{loc-alt-sig}) for $n\to\infty$
\begin{eqnarray*}
\frac{I_n(\hat\theta)-\mu}{\tau}&\overset{\mathcal{D}}{\longrightarrow}&\mathcal{N}(0,1)
\end{eqnarray*}
with 
\begin{eqnarray*}
\mu&=&E\left[\int d(w,V_1)d(w,V_2)f_W^2(w)f_{W|V}(w|V_1)f_{W|V}(w|V_2)\, dw\,\psi(V_1-V_2)\right]\\
\tau^2&=&2\sigma^4E\left[\int f_W^4(w)f_{W|V}(w|V_1)f_{W|V}(w|V_2)\, dw\,\psi^2(V_1-V_2)\right]\int K^2(u)\,du.
\end{eqnarray*}
\end{cor}

The asymptotic variance can be consistently estimated by 
	\begin{align*}
	\hat{\tau}^2&=\frac{2h^p}{n^6}{\sum_{i,j,k,l,k',l'}}^{\neq}(\Lambda_{\hat{\theta}}(Y_i)-\Lambda_{\hat{\theta}}(Y_k))(\Lambda_{\hat{\theta}}(Y_i)-\Lambda_{\hat{\theta}}(Y_{k'}))(\Lambda_{\hat{\theta}}(Y_j)-\Lambda_{\hat{\theta}}(Y_l))(\Lambda_{\hat{\theta}}(Y_j)-\Lambda_{\hat{\theta}}(Y_{l'}))
	\\[0,2cm]&\quad\qquad \times L_g(W_i-W_k)L_g(W_i-W_{k'})L_g(W_j-W_l)L_g(W_j-W_{l'})\big(K_h(W_i-W_j)\big)^2\psi^2(V_i-V_j)
	\end{align*}
in order to apply asymptotic tests. 
In contrast to the empirical characteristic function approach by Allison et al.\ (2018) the asymptotic distribution of the test we suggest is not influenced by the pre-selection of the transformation. 
That means that the test by Lavergne et al.\ (2015) (either using the asymptotic normality or a standard wild bootstrap) can simply be applied to the randomly transformed data. 
In the next section we consider bootstrap versions of the test. 

\begin{rem}
Further approaches for testing significance could be followed, e.g.\ the marked empirical process approach by Delgado and Gonzßlez-Manteiga (2001). However, we conjecture that for this approach the estimation of the transformation has to be taken into account in the asymptotic distribution and the bootstrap procedure. 
\end{rem}

\subsection{Bootstrap versions}\label{significance-bootstrap}

The standard  wild bootstrap (swb) versions of the test statistic $I_n(\hat\theta)$ is
\begin{eqnarray*}
{I}_n^{\rm swb*}&= &\frac{h^{\frac{p}{2}}}{n^3}{\sum_{i,j,k,l}}^{\neq}\left(Z_i^*-Z_k^*\right)\left(Z_j^*-Z_l^*\right)L_g(W_i-W_k)L_g(W_j-W_l)
	K_h(W_i-W_j)\psi(V_i-V_j)
\end{eqnarray*}
where we use the notations $Z_i=\Lambda_{\hat\theta}(Y_i)$ and $Z_i^*=\hat m_0(W_i)+U_i(Z_i-\hat m(X_i))$, $i=1,\dots,n$. Here, $\hat m_0$ is the Nadaraya-Watson estimator based on the sample $(W_i,Z_i)$, $i=1,\dots,n$, with kernel $L$ and bandwidth $g$, while $\hat m$ is a nonparametric estimator for $m(\cdot)=E[\Lambda_{\theta_0}(Y_1)| X_1=\cdot]$ based on $(X_i,Z_i)$, $i=1,\dots,n$. Further, $U_1,\dots,U_n$ are as in section \ref{bootstrapversions}.
The transformation wild bootstrap (twb) version of the test statistic $I_n(\hat\theta)$ is
\begin{eqnarray*}
{I}_n^{\rm twb*}&= &\frac{h^{\frac{p}{2}}}{n^3}{\sum_{i,j,k,l}}^{\neq}\left(\Lambda_{\hat\theta^*}(Y_i^*)-\Lambda_{\hat\theta^*}(Y_k^*)\right)
\left(\Lambda_{\hat\theta^*}(Y_j^*)-\Lambda_{\hat\theta^*}(Y_l^*)\right)L_g(W_i-W_k)L_g(W_j-W_l)
	\\
&&\qquad{}\times K_h(W_i-W_j)\psi(V_i-V_j),
\end{eqnarray*}
where $Y_i^*=\Lambda_{\hat\theta}^{-1}(\hat m_0(W_i)+\varepsilon_i^*)$ and $\varepsilon_i^*=\zeta_i^*+a_n\xi_i^*$ with $\zeta_i^*$ drawn with replacement from $\{\Lambda_{\hat\theta}(Y_j)-\hat m(X_j)\mid j=1,\dots,n\}$, $i=1,\dots,n$. Further, $\xi_1,\dots,\xi_n$ and $a_n$ are as in section \ref{bootstrapversions}, and $\hat\theta^*$ is the transformation parameter estimator built from $(X_i,Y_i^*)$, $i=1,\dots,n$. 
For the multiplier bootstrap versions note that ${I}_n(\hat\theta)$ is asymptotically equivalent to 
\begin{eqnarray*}
\tilde{I}_n(\hat\theta)&=&
\frac{h^{\frac{p}{2}}}{n}{\sum_{i,j}}^{\neq}K_h(W_i-W_j)\psi(V_i-V_j)\hat f_W(W_i)\hat f_W(W_j)\hat e_i\hat e_j
\end{eqnarray*}
where $ \hat e_i= \Lambda_{\hat\theta}(Y_i)-\hat m_0(W_i)$, $i=1,\dots,n$, and $\hat f_W$ is the kernel density estimator with kernel $L$ and bandwidth $g$ based on $W_1,\dots,W_n$. Then the multiplier bootstrap (mb) test statistic and its `centered' version (cmb) are
\begin{eqnarray*}
{I}_n^{\rm mb*}&=&
\frac{h^{\frac{p}{2}}}{n}{\sum_{i,j}}^{\neq}K_h(W_i-W_j)\psi(V_i-V_j)\hat f_W(W_i)\hat f_W(W_j)\hat e_i\hat e_j\xi_i\xi_j\\
{I}_n^{\rm cmb*}&=&
\frac{h^{\frac{p}{2}}}{n}{\sum_{i,j}}^{\neq}K_h(W_i-W_j)\psi(V_i-V_j)\hat f_W(W_i)\hat f_W(W_j)(\hat e_i\xi_i-\overline{\hat e\xi}_n)(\hat e_i\xi_j-\overline{\hat e\xi}_n)
\end{eqnarray*}
with $\xi_1,\dots,\xi_n$ and notations as in section \ref{bootstrapversions}. 

Approximation of the critical values of $I_n(\hat\theta)$ from either of these four bootstrap versions leads to consistent asymptotic level-$\alpha$ tests.

\section{Finite sample properties}\label{simus}
\def\theequation{5.\arabic{equation}}
\setcounter{equation}{0}

In this section we perform some simulations in order to compare the asymptotic versions of the proposed tests with the different bootstrap approaches and with competing procedures from the literature. In both of the following subsections we try to mimic the simulation settings of, on the one hand, Colling and Van Keilegom (2017) and, on the other hand, Allison et al.\ (2018). We use the language R (R Core Team, 2013) for our simulations. Throughout this section we consider the Yeo-Johnson transformation (see Yeo and Johnson, 2000)
$$\Lambda_{\theta}(Y)=\left\{\begin{array}{rc}\frac{(Y+1)^{\theta}-1}{\theta},&\textup{if }Y\geq0,\theta\neq0\\
\log(Y+1),&\textup{if }Y\geq0,\theta=0\\
-\frac{(1-Y)^{2-\theta}-1}{2-\theta},&\textup{if }Y<0,\theta\neq2\\
-\log(1-Y),&\textup{if }Y<0,\theta= 2.\end{array}\right.$$

\subsection{Lack-of-Fit Testing}\label{simulationlackoffit}

Let $m_{\theta}(x)=E[\Lambda_{\theta}(Y)|X=x]$ denote the expectation of $\Lambda_{\theta}(Y)$ conditioned on $X=x$. Like Colling and Van Keilegom (2017) we use the profile-likelihood estimator developed by Linton et al.\ (2008) for our simulations, that is
$$\hat{\theta}=\arg\,\underset{\theta}{\max}\,\sum_{i=1}^n\bigg(\log\hat{f}_{\varepsilon(\theta)}(\Lambda_{\theta}(Y_i)-\hat{m}_{\theta}(X_i))+\log\Lambda_{\theta}'(Y_i)\bigg),$$
where $\hat{f}_{\varepsilon(\theta)}$ is an estimator for the density of $\varepsilon(\theta)=\Lambda_\theta(Y)-m_\theta(X)$, $\hat{m}_{\theta}$ denotes an estimator for $m_{\theta}$, and $\Lambda'_\theta(y)=\partial \Lambda_\theta(y)/\partial y$. Here we implement $\hat{f}_{\varepsilon(\theta)}$ as an ordinary kernel estimator with Epanechnikov kernel and a bandwidth  following the normal reference rule. Further, we use a local linear estimator $\hat{m}_\theta$ (see Fan and Gijbels, 1996, for a detailed description) with a bandwidth obtained by cross validation. Note that the bandwidths also may depend on $\theta$. The estimator of $\theta$ is obtained iteratively with the function {\sl optimize} in R over the interval $[-1,2]$.
From now on we consider the model
$\Lambda_{\theta_0}(Y)=m(X)+\varepsilon$
with the null hypothesis
$$H_0: m\in\{m(\cdot,\beta):\beta\in B\}=\{x\mapsto\beta_1+\beta_2x:\beta_1,\beta_2\in\mathbb{R}\}$$
for the true regression parameter $\beta_0=(3,5)^t$. In order to examine the performance of the test under several alternatives we add the  deviation functions $\Delta(x)=2x^2,\Delta(x)=3x^2,\Delta(x)=4x^2,\Delta(x)=5x^2,\Delta(x)=2\exp(x),\Delta(x)=3\exp(x),\Delta(x)=4\exp(x),\Delta(x)=5\exp(x),\Delta(x)=0.25\sin(2\pi x),\Delta(x)=0.5\sin(2\pi x),\Delta(x)=0.75\sin(2\pi x)$, and $\Delta(x)=\sin(2\pi x)$. Here $X$ and $\varepsilon$ follow a uniform distribution on $[0,1]$ and a standard normal distribution truncated on $[-3,3]$, respectively.\\
First, tables \ref{table1}--\ref{table3} show the (empirical) rejection probabilities of the tests $T_n(\hat\theta)$ and $V_n(\hat\theta)$ developed in section \ref{section-gof} this paper, where $V_n(\hat{\t})$ is scaled with its estimated standard deviation. The simulations are conducted for a sample size of $n=200$ with $B=1000$ bootstrap repetitions for the standard wild bootstrap, the multiplier bootstrap and the centered multiplier bootstrap and $B=250$ bootstrap repetitions for the transformation wild bootstrap in each of $500$ simulation runs. Further, the transformation parameter is chosen to be equal to $\theta_0=0$ (table \ref{table1}), $\theta_0=0.5$ (table \ref{table2}) and $\theta_0=1$ (table \ref{table3}). The significance level is chosen to be $\alpha=0.10$. We show rejection probabilities based on the standard wild bootstrap (swb), the transformation wild bootstrap (twb), the multiplier bootstrap (mb), the centered multiplier bootstrap (cmb), and based on the asymptotic distribution (asym) as given in Corollary \ref{cor-gof}. For the twb, like Colling and Van Keilegom, we choose $a_n=0.1$.
 In the last two columns we show the rejection probabilities of the test based on the test statistic $W_{\exp}^2$ and the maximum of the rejection probabilities of the tests based on $T_{CM},W_{1}^2,W_{\exp_i}^2,W_{\exp}^2,W_{1/\exp}^2$ and $W_{\sin}^2$ from Colling and Van Keilegom (2017). Note that although in their article the test based on $W_{\exp}^2$ seems to perform best in most of the cases, this heavily depends on the alternative. For example, for the nonmonotone alternative $\sin(2\pi x)$, the performance of the test based on $W_{\exp}^2$ is rather poor. 
 Note also that the last column does not represent the power of an actual testing procedure but only serves as comparison. 
 
A conclusion of the comparison of the tests corresponding to $T_n(\hat{\theta})$ and $V_n(\hat{\theta})$ on the basis of the tables \ref{table1}--\ref{table3} is ambiguous. While for the swb and the twb approach the test related to that of Zheng performs better in most of the cases, the opposite is true for the mb and the cmb approach as well as the tests based on the asymptotic distribution. Further, for both test statistics and all three transformation parameters the level is slightly overestimated when applying swb or twb. Therefore, we would suggest using the cmb approach on the whole, since the level is well approximated  and the rejection probabilities under all alternatives are rather high. Indeed, as can be seen in tables \ref{table1}--\ref{table3}, if $T_n(\hat{\theta})$ is combined with the cmb, the resulting rejection probability in most of the cases is even higher than the maximum of those calculated using $T_{CM},W_{1}^2,W_{\exp_i}^2,W_{\exp}^2,W_{1/\exp}^2$ without the drawback of vanishing power under specific alternatives.
 In particular the immense computational costs of the transformation wild bootstrap is not justified. Another problem of the transformation wild bootstrap is the instability of the resulting procedure due to the additional estimation of the transformation parameter for every bootstrap sample, because when adding the bootstrap residuals to the estimated regression function in order to calculate $Y_i^*=\Lambda_{\hat{\theta}}^{-1}(m(X_i^*,\hat{\beta})+\varepsilon_i^*)$ it might happen that the argument of the inverse transformation function leaves its domain of definition. This has to be taken account of in the algorithm. 

\begin{center} TABLES \ref{table1}--\ref{table3} HERE\end{center}


\subsection{Testing Significance of Covariables}

In the second part of our simulation study, we examine the small sample size behaviour of the test developed in section \ref{testsignificance} in order to compare it to the transformation wild bootstrap test provided by Allison et al.\ (2018). We use the same simulation setting as in the reference.  The estimation of the transformation parameter is conducted as in the last subsection. Further, $(W,V)$ follows a uniform distribution on $[0,1]^2$, while the residuals are standard normally distributed. In our test statistic $I_n(\hat\theta)$  we use the Epanechnikov kernel for $K$ and $L$ with bandwidths $h$ and $g$ selected by cross validation, and density $\psi$ of a centered normal distribution with variance $v=0.10$. 
We consider the standard wild bootstrap (swb),  multiplier bootstrap (mb), centered multiplier bootstrap (cmb) and transformation wild bootstrap (twb) as in section \ref{significance-bootstrap}.  The simulations are conducted for the sample sizes $n=75$ and $n=100$ at the significance level $0.05$. We use $B=1000$ bootstrap data for the swb, the mb and the cmb, but only $B=250$ for the twb which is computationally demanding. 
The results are based on $500$ simulation runs. The models considered are
\begin{align}
&\Lambda_{\theta_0}(Y)=1+W+\varepsilon\label{null}\\[0,2cm]
&\Lambda_{\theta_0}(Y)=W+V+\varepsilon\label{V}\\[0,2cm]
&\Lambda_{\theta_0}(Y)=3+2W+0.25\sin(2\pi V)+\varepsilon\label{0.25sin(2piV)}\\[0,2cm]
&\Lambda_{\theta_0}(Y)=1+W+\sin(5V)+\varepsilon\label{sin(5V)}\\[0,2cm]
&\Lambda_{\theta_0}(Y)=3+2W+5V^2\varepsilon\label{5V2}\\[0,2cm]
&\Lambda_{\theta_0}(Y)=1+W+V^2+\varepsilon\label{V2}\\[0,2cm]
&\Lambda_{\theta_0}(Y)=1+W+\exp(V^2)+\varepsilon\label{exp(V2)}
\end{align}
for the true transformation parameter $\theta_0=1$.
The results depicted in table \ref{table5} show that for $I_n(\hat\theta)$ the mb and the twb have no advantage over the cmb and the swb. The swb performs slightly better than the cmb. Nevertheless, all four tests approximate the level quite well or are even a bit conservative.\\
We also compare our tests with those based on the test statistics $\psi_{n,1}$ and $\tilde{\Psi}_n$ from Allison et al.\ (2018) (for their choice of parameters $a=0.25$ and $\gamma=0.01$), as these tests are recommended in their paper. Apart from model (\ref{V}), the cmb and the swb approach outperform those by Allison et al.\ (2018), in some cases in fact with a substantial difference.\\
Additionally, tables \ref{table6} and \ref{table7} show for the mb and cmb approaches the influence of the choice of the bandwidth and the parameter $v$ of the  weight function $\psi$, which was chosen as density of the centered normal distribution with variance $v$.  In table \ref{table5}, a bandwidth obtained via cross validation (in the following called $cv$) was used. In table \ref{table6}, the rejection probabilities for $v=0.1$ and the four cases $h=cv,h=0.5cv,h=2cv,h=0.2$ as well as the mean of the $cv$-bandwidth are given. Although with a growing bandwidth the rejection probabilities seem to increase, all choices lead to reasonable tests which indicates that the test is rather stable with respect to the bandwidth. The same conclusion can be drawn when considering different values of $v\in \{0.05, 0.1, 0.2\}$. Even though the choice $v=0.1$ seems to do best, the difference regarding the rejection probabilities is rather small.

\begin{center} TABLES \ref{table5}--\ref{table7} HERE\end{center}


\begin{appendix}

\section{Assumptions}
\def\theequation{A.\arabic{equation}}
\setcounter{equation}{0}

\subsection{Assumptions for Theorems \ref{theo1} and \ref{theo-boot-gof}}

\begin{enumerate}[label=(\textbf{A\arabic{*}})]
	\item\label{A1} On its compact support $\mathcal{X}$ the density $f_X$ of the covariate $X$ is continuously differentiable.
	\item\label{A2} $\varepsilon$ is a centered, non-degenerate random variable with $E[\varepsilon^4]<\infty$.
	\item\label{A3} $K$ is a symmetric and continuously differentiable density with compact support.
	\item\label{A4} The bandwidth $h$ fulfills $h\rightarrow0$ and $nh^d\rightarrow\infty$.
	\item\label{A5} $\Delta_n$ is continuously differentiable and $\Delta_n$ and its derivative are bounded uniformly in $x$ and $n$. Further,  $c_n\sim n^{-\frac{1}{2}}h^{-\frac{d}{4}}$.
	\item\label{A6} $\hat{\theta}$ is an estimator of $\theta_0$ fulfilling $\hat{\theta}-\theta_0=O_P\big(n^{-\frac{1}{2}}\big)$.
	\item\label{A7} For all $y$ the transformation $\Lambda_{\theta}(y)$ is two times continuously differentiable with respect to $\theta$. We denote the gradient and Hessian matrix with $\dot{\Lambda}_{\theta}(y)$ and $\ddot{\Lambda}_{\theta}(y)$, respectively. Let
	$\sup_{x\in\mathcal{X}}E[\|\dot{\Lambda}_{\theta_0}(Y)\|^2|X=x]<\infty$
	and let there exist an $\alpha>0$ with
	$E[\sup_{\theta':\|\theta'-\theta_0\|\leq\alpha}\|\ddot\Lambda_{\theta'}(Y)\|]<\infty$.
	\item\label{A8} The map $x\mapsto E[\Lambda_{\theta_0}(Y)^4|X=x]$ is continuously differentiable with $E[E[\Lambda_{\theta_0}(Y)^4|X]^2]<\infty$.
	\item\label{A9} All partial derivatives of $m(x,\beta)$ of order $0,\dots,3$ with respect to $x$ and $\beta$ exist and are continuous for all $(x,\beta)$. The set $B$ of regression parameters $\beta$ is compact and
	$\bar{\beta}=\arg\min_{\beta\in B}\,E[(m(X)-m(X,\beta))^2]$
	is (for all $n$) a uniquely determined interior value of $B$ with $\|\bar{\beta}-\beta_0\|\rightarrow0$ for some inner point $\beta_0\in\mathring{B}$ as $n\to\infty$. 
	Moreover,
	$$\underset{\|\beta-\beta_0\|>\delta}{\inf}E[(m(X,\beta)-m(X,{\beta_0}))^2]>0 $$
	holds for all $\delta>0$ and $n$ sufficiently large. The matrix
$$\Omega(\beta_0)=\bigg(E\bigg[\frac{\partial}{\partial\beta_i}m(X,\beta)\bigg|_{\beta=\beta_0}\frac{\partial}{\partial\beta_j}m(X,\beta)\bigg|_{\beta=\beta_0}\bigg]\bigg)_{i,j=1,...,q}$$
	is regular.
\end{enumerate}

\begin{rem}\label{Delta-fixed} Under fixed alternatives (see Theorem \ref{theo-boot-gof}) $\Delta_n=\Delta$ in assumption \ref{A5} does not depend on $n$ and neither does $\bar\beta=\beta_0$ in assumption \ref{A9}. Under the null hypothesis, $\Delta_n=\Delta\equiv 0$, while $\bar\beta=\beta_0$ is the true regression parameter. 
\end{rem}

\subsection{Assumptions for Theorem \ref{theo2}}

	Denote with $\mathcal{U}^p$ the set of all integrable and uniformly continuous functions of $p$ variables. 
	\begin{enumerate}[label=(\textbf{B\arabic{*}})]
	\item\label{B1} The covariate $W$ has compact support $\mathcal{W}$ and density $f_W$.  On their support the functions $f_W$ and $r(\cdot)f_W(\cdot)$ lie in $\mathcal{U}^p$ and are twice continuously differentiable.
	\item\label{B2}  It holds that $E[|d(W,V)|]<\infty$,  $E[d(W,V)|W]=0$ and $\delta_n\sim n^{-1/2} h^{-p/4}$.
	\item\label{B3} For any $z$ in the support of $V$ the random variable $W$ admits a conditional density denoted by $f_{W|V}(\cdot|v)$. Further, $E[(\Lambda_{\theta_0}(Y)-r(W))^2|W=\cdot]f_W(\cdot)$ and $E[(\Lambda_{\theta_0}(Y)-r(W))^4|W=\cdot]f^4_W(\cdot)$ belong to $\mathcal{U}^p$.
	\item\label{B4} 
Let $K$ and $L$ be symmetric kernel functions of order 2 with compact support. Let $K$ have an almost everywhere positive and integrable Fourier
transform  and let $L$ be of bounded variation.
	\item\label{B5} Let $\sigma^2(w,v)=E[(\Lambda_{\theta_0}(Y)-r(W))^2|W=w,V=v]$, then for all $v$ in the support of $V$ the function $\sigma^2(\cdot,v)f^2_W(w)f_{W|V}(\cdot|v)\in\mathcal{U}^p$ has an integrable Fourier transform and $E[\sigma^4(W,X)f_W^4(W)f_{W|V}(W|X)]<\infty$.
	\item\label{B6} $E[d^2(W,V)|W=\cdot]f_W^2(\cdot)$ belongs to $\mathcal{U}^p$, $d(\cdot,x)f_W(\cdot)f_{W|V}(\cdot|x)$ is integrable and square integrable for any $x$ in the support of $X$. Further, $E[d^2(W,X)f^2_W(W)f_{W|V}(W|X)]<\infty$.
	\item\label{B7} $\psi$ is a bounded function with positive and integrable Fourier transform.
	\item\label{B8} The bandwidths $g$ and $h$ fulfill
	$$g\rightarrow0,\quad h\rightarrow0,\quad \frac{ng^{\frac{4p}{3}}}{\log n}\rightarrow\infty,\quad nh^p\rightarrow\infty, \quad nh^{\frac{p}{2}}g^{4}\rightarrow0, \quad \frac{h}{g}\rightarrow0.$$
	\item\label{B9} It holds that $E[\Lambda_{\theta_0}(Y)^8]<\infty$.
	\item\label{B10} $\hat{\theta}$ is an estimator of $\theta_0$ fulfilling $\hat{\theta}-\theta_0=O_P\big(n^{-\frac{1}{2}}\big)$.
	\item\label{B11} For all $y$ the transformation $\Lambda_{\theta}(y)$ is two times continuously differentiable with respect to $\theta$. Further,
	$\sup_{w\in\mathcal{W}}E[\|\dot{\Lambda}_{\theta_0}(Y)\|^2|W=w]<\infty$
	holds and there exists an $\alpha>0$ with
	$E[\sup_{\theta':\|\theta'-\theta_0\|\leq\alpha}\|\ddot\Lambda_{\theta'}(Y)\|]<\infty$.
	\end{enumerate} 

	\begin{rem}
		Note that $E[\varepsilon^8]<\infty$ is implied by \ref{B9}.
		Instead of $nh^{\frac{p}{2}}g^{4}\rightarrow0$ Lavergne et al.\ (2015) use the assumption $nh^{\frac{p}{2}}g^{2s}\rightarrow0$ for some $s\geq2$ with some additional differentiability conditions on $f_W$ and $r(\cdot)f_W(\cdot)$ and a higher order kernel for $L$. The stronger version arises here due to boundary problems.
		The reason for considering a density with a compact support is to allow for example the profile likelihood estimator of Linton et al.\ (2008) and other estimators for $\theta_0$, that require a compact support of the covariate. By straightforward calculations it can be seen that assumption \ref{B8} can be fulfilled if and only if $p\leq5$.
	\end{rem}

\section{Proofs}
\def\theequation{B.\arabic{equation}}
\setcounter{equation}{0}

\subsection{Proof of Theorem \ref{theo1}}\label{prooftheo1}

\subsubsection*{Proof for $T_n$.}

Consider the decomposition $T_n(\hat\theta)=T_n(\theta_0)+I+II+III+IV+V$, where
\begin{eqnarray*}
I &=& nh^{\frac{d}{2}}\int\bigg(\frac{1}{n}\sum_{i=1}^nK_h(x-X_i)(\Lambda_{\hat{\theta}}(Y_i)-\Lambda_{\theta_0}(Y_i))\bigg)^2\,dx\\
II &=& nh^{\frac{d}{2}}\int\bigg(\frac{1}{n}\sum_{i=1}^nK_h(x-X_i)(m(X_i,{\hat{\beta}_{\theta_0}})-m(X_i,{\hat{\beta}_{\hat{\theta}}}))\bigg)^2\,dx\\
III &=& 2nh^{\frac{d}{2}}\int\bigg(\frac{1}{n}\sum_{i=1}^nK_h(x-X_i)(\Lambda_{\theta_0}(Y_i)-m(X_i,{\hat{\beta}_{\theta_0}}))\bigg)
\\&&\qquad\quad\times\bigg(\frac{1}{n}\sum_{i=1}^nK_h(x-X_i)(\Lambda_{\hat{\theta}}(Y_i)-\Lambda_{\theta_0}(Y_i))\bigg)\,dx\\
IV &=& 2nh^{\frac{d}{2}}\int\bigg(\frac{1}{n}\sum_{i=1}^nK_h(x-X_i)(\Lambda_{\theta_0}(Y_i)-m(X_i,{\hat{\beta}_{\theta_0}}))\bigg)\\
&&\qquad\quad\times\bigg(\frac{1}{n}\sum_{i=1}^nK_h(x-X_i)(m(X_i,{\hat{\beta}_{\theta_0}})-m(X_i,{\hat{\beta}_{\hat{\theta}}}))\bigg)\,dx\\
V&=& 2nh^{\frac{d}{2}}\int\bigg(\frac{1}{n}\sum_{i=1}^nK_h(x-X_i)(\Lambda_{\hat{\theta}}(Y_i)-\Lambda_{\theta_0}(Y_i))\bigg)\\
&&\qquad\quad\times\bigg(\frac{1}{n}\sum_{i=1}^nK_h(x-X_i)(m(X_i,{\hat{\beta}_{\theta_0}})-m(X_i,{\hat{\beta}_{\hat{\theta}}}))\bigg)\,dx.
\end{eqnarray*}
In what follows we show asymptotic negligibility of the terms $I,II, III, IV$ and $V$. 

To treat $I$ we will use a Taylor expansion of $\Lambda_{\hat\theta}(Y_i)$ around $\Lambda_{\theta_0}(Y_i)$. To this end note that due to assumption \ref{A6} we can assume $\|\hat\theta-\theta_0\|\leq\alpha$ with $\alpha$ from assumption \ref{A7}. Let now $\theta_i^*$ denote suitable parameters between $\hat\theta$ and $\theta_0$ such that 
\begin{eqnarray}
\nonumber
I&=&\frac{h^{\frac{d}{2}}}{n}\sum_{i=1}^n\sum_{j=1}^n\int K_h(x-X_i)(\dot{\Lambda}_{\theta_0}(Y_i)(\hat{\theta}-\theta_0)+(\hat{\theta}-\theta_0)^t\ddot\Lambda_{\theta^*_i}(Y_i)(\hat{\theta}-\theta_0))
\\
\nonumber
&&\qquad\qquad {}\times K_h(x-X_j)(\dot{\Lambda}_{\theta_0}(Y_j)(\hat{\theta}-\theta_0)+(\hat{\theta}-\theta_0)^t\ddot\Lambda_{\theta^*_j}(Y_j)(\hat{\theta}-\theta_0))\,dx
\\
\label{I1}
&\leq&\frac{h^{\frac{d}{2}}\|\hat{\theta}-\theta_0\|^2}{n}\sum_{i=1}^n\sum_{j=1}^n\|\dot{\Lambda}_{\theta_0}(Y_i)\|\|\dot{\Lambda}_{\theta_0}(Y_j)\|\int K_h(x-X_i)K_h(x-X_j)\,dx
\\
\label{I2}
&&{}+\frac{2h^{\frac{d}{2}}\|\hat{\theta}-\theta_0\|^3}{n}\sum_{i=1}^n\sum_{j=1}^n\|\dot{\Lambda}_{\theta_0}(Y_i)\|\|\ddot\Lambda_{\theta^*_j}(Y_j)\|\int K_h(x-X_i)K_h(x-X_j)\,dx
\\
\label{I3}
&&{}+\frac{h^{\frac{d}{2}}\|\hat{\theta}-\theta_0\|^4}{n}\sum_{i=1}^n\sum_{j=1}^n\|\ddot\Lambda_{\theta^*_i}(Y_i)\|\|\ddot\Lambda_{\theta^*_j}(Y_j)\|\int K_h(x-X_i)K_h(x-X_j)\,dx.
\end{eqnarray}
For the expectation of the summands of term (\ref{I1}) we have for $i=j$
\begin{align*}
&E\left[\|\dot{\Lambda}_{\theta_0}(Y_i)\|^2\int K_h(x-X_i)^2\,dx\right] \leq Ch^{-d} E[\|\dot{\Lambda}_{\theta_0}(Y_i)\|^2]= O(h^{-d})
\end{align*}
and for $i\neq j$
\begin{align*}
&E\bigg[\|\dot{\Lambda}_{\theta_0}(Y_1)\|\|\dot{\Lambda}_{\theta_0}(Y_2)\|\int K_h(x-X_1)K_h(x-X_2)\,dx\bigg]
\\[0,2cm]&=\int \big(E\big[\|\dot{\Lambda}_{\theta_0}(Y_1)\|K_h(x-X_1)\big]\big)^2\,dx
\\[0,2cm]&=\int \Big(E\Big[E\big[\|\dot{\Lambda}_{\theta_0}(Y_1)\|\,\big|X_1\big]K_h(x-X_1)\Big]\Big)^2\,dx \leq C \int \Big(E\Big[K_h(x-X_1)\Big]\Big)^2\,dx =O(1),
\end{align*}
where the second last equality follows by assumption \ref{A7} for some constant $C$.
Thus, term (\ref{I1}) is, due to assumption \ref{A6}, of order $O_P(h^{d/2}n^{-2})(O_P((h^{-d}n)+O_P(n^2))=o_P(1)$. 

For term (\ref{I2}) one obtains
\begin{align*}
&O_P(h^{{d/2}}n^{-5/2})\sum_{i=1}^n\sum_{j=1}^n\|\dot{\Lambda}_{\theta_0}(Y_i)\|\|\ddot\Lambda_{\theta^*_j}(Y_j)\|\int K(x)K_h(hx+X_i-X_j)\,dx
\\[0,2cm]&\leq O_P(h^{-{d/2}}n^{-1/2})\bigg(\frac{1}{n}\sum_{i=1}^n\|\dot{\Lambda}_{\theta_0}(Y_i)\|\bigg)\bigg(\frac{1}{n}\sum_{j=1}^n\sup_{\theta:\|\theta-\theta_0\|\leq\alpha}\|\ddot\Lambda_{\theta}(Y_j)\|\bigg)=o_P(1)
\end{align*}
by assumption \ref{A7}. The last term (\ref{I3}) is treated similarly and we obtain $I=o_P(1)$. 

Term $II$ is treated completely analogously with a Taylor expansion of $m(X_i,\hat\beta_{\hat\theta})$ around $m(X_i,\hat\beta_{\theta_0})$ and using $\hat\beta_{\hat\theta}-\hat\beta_{\theta_0}=O_P(n^{-1/2})$. For the latter equality note that $\hat\beta_{\hat\theta}-\bar\beta=O_P(n^{-1/2})$ follows from the proof of Lemma 4 in Colling and Van Keilegom (2016) (taking into account that our local alternative has a slower rate), whereas $\hat\beta_{\theta_0}-\bar\beta=O_P(n^{-1/2})$ follows from standard arguments in the model $\Lambda_{\theta_0}(Y)=m(X)+\varepsilon$ with known transformation. 

With the Cauchy-Schwarz inequality $V=o_P(1)$ directly follows from $I=o_P(1)$ and $II=o_P(1)$. 

We further decompose the term $III=A_n+B_n+C_n$ into the terms
\begin{eqnarray*}
A_n&=& 2nh^{\frac{d}{2}}\int\bigg(\frac{1}{n}\sum_{i=1}^nK_h(x-X_i)(m(X_i,\bar{\beta})-m(X_i,{\hat{\beta}_{\theta_0}}))\bigg)\\
&&\quad\qquad\times\bigg(\frac{1}{n}\sum_{j=1}^nK_h(x-X_j)(\Lambda_{\hat{\theta}}(Y_j)-\Lambda_{\theta_0}(Y_j))\bigg)\,dx\\
B_n&=&2nh^{\frac{d}{2}}\int\bigg(\frac{1}{n}\sum_{i=1}^nK_h(x-X_i)(\Lambda_{\theta_0}(Y_i)-m(X_i))\bigg)\\
&&\quad\qquad\times\bigg(\frac{1}{n}\sum_{j=1}^nK_h(x-X_j)(\Lambda_{\hat{\theta}}(Y_j)-\Lambda_{\theta_0}(Y_j))\bigg)\,dx\\
C_n&=&2nh^{\frac{d}{2}}\int\bigg(\frac{1}{n}\sum_{i=1}^nK_h(x-X_i)(m(X_i)-m(X_i,\bar{\beta}))\bigg)\\
&&\quad\qquad\times\bigg(\frac{1}{n}\sum_{j=1}^nK_h(x-X_j)(\Lambda_{\hat{\theta}}(Y_j)-\Lambda_{\theta_0}(Y_j))\bigg)\,dx.
\end{eqnarray*} 
Now $A_n$ is very similar to $V$ and can be treated in the same way noting that $\bar{\beta}-\hat{\beta}_{\theta_0}=O_P(n^{-1/2})$. 

With Taylor expansion we obtain for some $\theta^*_j$ between $\hat\theta$ and $\theta_0$,
\begin{eqnarray*}
B_n&=&
2\frac{h^{\frac{d}{2}}}{n}\sum_{i=1}^n\sum_{\stackrel{\scriptstyle j=1}{j\neq i}}^n\int K_h(x-X_i)\varepsilon_iK_h(x-X_j)\,dx\dot{\Lambda}_{\theta_0}(Y_j)(\hat{\theta}-\theta_0)
\\
&&{}+2\frac{h^{\frac{d}{2}}}{n}\sum_{i=1}^n\int K_h(x-X_i)^2\varepsilon_i\dot{\Lambda}_{\theta_0}(Y_i)\,dx\,(\hat{\theta}-\theta_0)
\\
&&{}+2\frac{h^{\frac{d}{2}}}{n}\sum_{i=1}^n\sum_{j=1}^n\int K_h(x-X_i)K_h(x-X_j)\varepsilon_i(\hat{\theta}-\theta_0)^t\ddot\Lambda_{\theta^*_j}(Y_j)(\hat{\theta}-\theta_0)\,dx.
\end{eqnarray*} 
The expectation of the absolute value of the integral in the second term can be bounded by 
\begin{align*}
&\quad\quad\  E\bigg[E[|\varepsilon_i|\|\dot{\Lambda}_{\theta_0}(Y_i)\|\,|X_i]\int K_h(x-X_i)^2\,dx\bigg]
\\
&\quad\quad\ \leq \sigma E\bigg[\sqrt{E[\|\dot{\Lambda}_{\theta_0}(Y_i)\|^2\,|X_i]}\int K_h(x-X_i)^2\,dx\bigg] =O(h^d)
\end{align*}
by assumption \ref{A7}. Thus, the second term in the decomposition of $B_n$ is of order \linebreak $O_P(h^{-d/2}n^{-1/2})=o_P(1)$. The third term can be bounded by 
$$2h^{\frac{d}{2}}\|\hat{\theta}-\theta_0\|^2\sum_{j=1}^n\|\ddot\Lambda_{\theta^*_j}(Y_j)\|\int K_h(x-X_j)\,dx\,\sup_{x}\Big|\frac{1}{n}\sum_{i=1}^n\varepsilon_iK_h(x-X_i)\Big|=o_P(h^{d/2})$$
(consider the last factor as local constant estimator in a regression model with zero regression function). 
For the first term in the decomposition of $B_n$ consider the double sum without the factor $\hat{\theta}-\theta_0=O_P(n^{-1/2})$. Because $\varepsilon$ is centered and independent of $X$, the expectation is zero and the variance is easy to derive. With calculations similar to before it can be shown that the variance is of order $O(nh^d)$ such that for the first term in $B_n$ one has $O_P((nh^d)^{1/2}n^{-1/2})=o_P(1)$.  

Inserting $m(X_i)-m(X_i,\bar\beta)=c_n\Delta_n(X_i)$ and the Taylor expansion for $\Lambda_{\hat{\theta}}(Y_j)-\Lambda_{\theta_0}(Y_j)$ one sees that $C_n$ has the rate 
$nh^{d/2}c_nO_P(\|\hat\theta-\theta_0\|)=O_P(h^{d/4})=o_P(1)$. 

The remaining term $IV$ is treated analogously to before inserting Taylor expansions of $\Lambda_{\hat{\theta}}(Y_i)$ around $\Lambda_{\theta_0}(Y_i)$ and of $m(X_i,{\hat{\beta}_{\hat{\theta}}})$ around $m(X_i,{\hat{\beta}_{\theta_0}})$. 
 \hfill $\Box$

\subsubsection*{Proof for $V_n$.}

We decompose $nh^{\frac{d}{2}}V_n(\hat\theta)=nh^{\frac{d}{2}}V_n(\theta_0)+\tilde I+\tilde{II} +\tilde{III}+\tilde{IV}+\tilde V$, where
\begin{eqnarray*}
\tilde{I} &=&\frac{h^{\frac{d}{2}}}{n-1}\sum_{i=1}^n\sum_{\stackrel{\scriptstyle j=1}{j\neq i}}^nK_h(X_i-X_j)(\Lambda_{\hat{\theta}}(Y_i)-\Lambda_{\theta_0}(Y_i))(\Lambda_{\hat{\theta}}(Y_j)-\Lambda_{\theta_0}(Y_j))
\\
\tilde{II} &=& \frac{h^{\frac{d}{2}}}{n-1}\sum_{i=1}^n\sum_{\stackrel{\scriptstyle j=1}{j\neq i}}^nK_h(X_i-X_j)
(m(X_i,{\hat{\beta}_{\theta_0}})-m(X_i,{\hat{\beta}_{\hat{\theta}}}))(m(X_j,\hat{\beta}_{\theta_0})-m(X_j,\hat{\beta}_{\hat{\theta}}))\\
\tilde{III} &=& \frac{2h^{\frac{d}{2}}}{n-1}\sum_{i=1}^n\sum_{\stackrel{\scriptstyle j=1}{j\neq i}}^nK_h(X_i-X_j)(\Lambda_{\theta_0}(Y_i)-m(X_i,{\hat{\beta}_{\theta_0}}))(\Lambda_{\hat{\theta}}(Y_j)-\Lambda_{\theta_0}(Y_j))\\
\tilde{IV} &=& \frac{2h^{\frac{d}{2}}}{n-1}\sum_{i=1}^n\sum_{\stackrel{\scriptstyle j=1}{j\neq i}}^nK_h(X_i-X_j)
(\Lambda_{\theta_0}(Y_i)-m(X_i,{\hat{\beta}_{\theta_0}}))(m(X_j,\hat{\beta}_{\theta_0})-m(X_j,\hat{\beta}_{\hat{\theta}}))\\
\tilde{V}&=& \frac{2h^{\frac{d}{2}}}{n-1}\sum_{i=1}^n\sum_{\stackrel{\scriptstyle j=1}{j\neq i}}^nK_h(X_i-X_j)(\Lambda_{\hat{\theta}}(Y_i)-\Lambda_{\theta_0}(Y_i))(m(X_j,\hat{\beta}_{\theta_0})-m(X_j,\hat{\beta}_{\hat{\theta}})),
\end{eqnarray*}
and we have to show that $\tilde I,\dots,\tilde V$ are of order $o_P(1)$. Note that rewriting the terms $I,\dots,V$ in the decomposition of $T_n(\hat\theta)$ one sees that those have the very same structure as the terms $\tilde I,\dots,\tilde V$. The only difference is that for the latter terms the double sums are without diagonal terms and the kernel $K_h(X_i-X_j)$ in  $\tilde I,\dots,\tilde V$ corresponds to $\int K_h(x-X_i)K_h(x-X_j)\,dx = (K*K)_h(X_i-X_j)$ in $I,\dots,V$.  The derivations are thus analogous and omitted for the sake of brevity. 
\hfill $\Box$

\subsection{Proof of Theorem \ref{theo-boot-gof}} \label{beweis-boot}

We sketch the proof for the multiplier bootstrap versions of  $T_n$; proofs for the versions of $V_n$ are analogous. One can show similarly to Colling and Van Keilegom (2016) that $\hat{\beta}_{\hat{\theta}}-\bar{\beta}=O_P(n^{-\frac{1}{2}})$ holds.
With $P^*$ we denote the conditional probability, given the sample $(X_1,Y_1),(X_2,Y_2),\dots$, and with $E^*$, $Var^*$ the corresponding conditional expectation and conditional variance, respectively.  

\subsubsection*{Proof for \rm $T_n^{\text{mb}*}$\bf.}

Write $T_n^{\text{mb}*}=\tilde{T}_{n,1}^{\text{mb}*}+\tilde{T}_{n,2}^{\text{mb}*}$ with 
\begin{eqnarray*}
\tilde{T}_{n,1}^{\text{mb}*} &=& \frac{h^{\frac{d}{2}}}{n}\sum_{i=1}^n(K\ast K)_h(0)\hat e_i^2\xi_i^2\\
\tilde{T}_{n,2}^{\text{mb}*} &=& \frac{h^{\frac{d}{2}}}{n}\sum_{i=1}^n\sum_{\overset{\scriptstyle j=1}{j\neq i}}^n(K\ast K)_h(X_i-X_j)\hat{e}_i\hat{e}_j\xi_i\xi_j.
\end{eqnarray*}
Recall the definition $\hat e_i= \Lambda_{\hat{\theta}}(Y_i)-m(X_i,{\hat{\beta}_{\hat{\theta}}})$ and define $e_i=\Lambda_{\theta_0}(Y_i)-m(X_i,\bar\beta)=\varepsilon_i+\Delta(X_i)$, $i=1,\dots,n$.
 Then, by the same reasoning as in section \ref{prooftheo1}, $\tilde{T}_{n,1}^{\text{mb}*}$ and $\tilde{T}_{n,2}^{\text{mb}*}$ can be asymptotically replaced by
\begin{eqnarray*}
	T_{n,1}^{\text{mb}*} &=& \frac{h^{\frac{d}{2}}}{n}\sum_{i=1}^n(K\ast K)_h(0)e_i^2\xi_i^2\\
	T_{n,2}^{\text{mb}*} &=& \frac{h^{\frac{d}{2}}}{n}\sum_{i=1}^n\sum_{\overset{\scriptstyle j=1}{j\neq i}}^n(K\ast K)_h(X_i-X_j)e_ie_j\xi_i\xi_j
\;=\; \sum_{i=1}^nZ_{n,i}^*,
\end{eqnarray*}
where $Z_{n,i}^*=2e_i\xi_ih^{d/2}n^{-1}\sum_{j=1}^{i-1}(K\ast K)_h(X_i-X_j)e_j\xi_j$.\\
Due to
\begin{eqnarray*}
T_n^{\text{cmb}*}&=&\frac{h^{\frac{d}{2}}}{n}\sum_{i=1}^n\sum_{j=1}^n(K\ast K)_h(X_i-X_j)\bigg(\hat{e}_i\xi_i-\frac{1}{n}\sum_{k=1}^n\hat{e}_k\xi_k\bigg)\bigg(\hat{e}_j\xi_j-\frac{1}{n}\sum_{k=1}^n\hat{e}_k\xi_k\bigg)
\\&=&T_n^{\text{mb}*}-\bigg(\frac{1}{n}\sum_{k=1}^n\hat{e}_k\xi_k\bigg)\frac{h^{\frac{d}{2}}}{n}\sum_{i=1}^n\sum_{j=1}^n(K\ast K)_h(X_i-X_j)(\hat{e}_i\xi_i+\hat{e}_j\xi_j)
\\&&\quad+\bigg(\frac{1}{n}\sum_{k=1}^n\hat{e}_k\xi_k\bigg)^2\frac{h^{\frac{d}{2}}}{n}\sum_{i=1}^n\sum_{j=1}^n(K\ast K)_h(X_i-X_j)
\end{eqnarray*}
and
$$E^*\bigg[\bigg(\frac{1}{n}\sum_{k=1}^n\hat{e}_k\xi_k\bigg)^2\bigg]=\frac{\sigma^2+E[\Delta(X_1)^2]}{n}+o_P\bigg(\frac{1}{n}\bigg),$$
it also holds that $T_n^{\text{cmb}*}$ is asymptotically equivalent to ${T}_{n,1}^{\text{mb}*}+{T}_{n,2}^{\text{mb}*}$ in terms of conditional convergence in probability. 
Straightforward calculations lead to
$$E[T_{n,1}^{\text{mb}*}]=h^{-\frac{d}{2}}E[e_i^2]\int K(x)^2\,dx\quad\textup{and}\quad Var(T_{n,1}^{\text{mb}*})=o(1),$$
so that $T_{n,1}^{\text{mb}*}=b_h^*+o_P(1)$.\\ 
For the treatment of $T_{n,2}^{\text{mb}*}$ note that $\{S_{n,k}^* =\sum_{i=1}^k Z_{n,i}^*\mid k=1,\dots,n\}$ is a martingale with respect to $P^*$ and the filtration $\{\mathcal{F}_{n,k}=\sigma(\xi_1,\dots,\xi_k)\mid k=1,\dots,n\}$ (for each $n$) and that $S_{n,n}^*=T_{n,2}^{\text{mb}*}$.
The assertion of  Theorem \ref{theo-boot-gof} follows if we show  $E^*[\exp(\text{i}t T_{n,2}^{\text{mb}*} )]=\exp(-(V^*)^2t^2/2)+o_P(1)$ for all $t\in\mathbb{R}$. To this end we will proceed in the same way as in Theorem 3.2 in Hall and Heyde (1980). For that purpose, we in turn apply the following auxiliary lemma which can be seen as the counterpart of Lemma 3.1 in Hall and Heyde (1980).

\begin{lemma}\label{lemmahallheyde}
	Under assumptions \ref{A1}--\ref{A9} it is
	\begin{equation}
	\sum_{i=1}^n{Z_{n,i}^*}^2=V^*+o_P(1)\quad\textup{and}\quad E^*\big[\underset{i=1,...,n}{\max}\,|Z_{n,i}^*|^4\big]=o_P(1).
	\end{equation}
\end{lemma}

\textbf{Proof.} We have $\sum_{i=1}^n{Z_{n,i}^*}^2=A_n^*+B_n^*$ with 
\begin{eqnarray*}
A_n^*&=&\frac{4h^d}{n^2}\sum_{i=1}^n\sum_{j=1}^{i-1}(K\ast K)_h(X_i-X_j)^2e_i^2e_j^2\xi_i^2\xi_j^2
\\
B_n^*&=&\frac{4h^d}{n^2}\sum_{i=1}^n\sum_{j=1}^{i-1}\sum_{\overset{\scriptstyle k=1}{k\neq j}}^{i-1}(K\ast K)_h(X_i-X_j)(K\ast K)_h(X_i-X_k)e_i^2e_je_k\xi_i^2\xi_j\xi_k
.
\end{eqnarray*}
Again, by the same reasoning as in section \ref{prooftheo1} it can be shown that
$$E[A_n^*]=V^*+o(1),\quad Var(A_n^*)=o(1),\quad E[B_n^*]=0,\quad\textup{and}\quad Var(B_n^*)=o(1),$$
which implies the first part of the assertion.
For the second part note that 
$$E^*\big[\underset{i=1,...,n}{\max}\,|Z_{n,i}^*|^4\big]\leq\sum_{i=1}^nE^*\bigg[\bigg(\frac{2h^{\frac{d}{2}}}{n}\sum_{j=1}^{i-1}(K\ast K)_h(X_i-X_j)e_ie_j\xi_i\xi_j\bigg)^4\bigg]=o_P(1),$$
where the last equality follows similarly to the considerations before. 
\hfill $\Box$

\medskip

As a direct consequence of Lemma \ref{lemmahallheyde} and the Cauchy-Schwarz inequality one has 
\begin{equation}\label{csu}
E^*[\underset{i=1,...,n}{\max}\,|Z_{n,i}^*|]=o_P(1)\quad\textup{and}\quad E^*[\underset{i=1,...,n}{\max}\,|Z_{n,i}^*|^2]=o_P(1).
\end{equation}
The basic difference between our arguments and those in Hall and Heyde (1980) consists in replacing the expectations by conditional expectations $E^*$. In what follows let $I\{\dots\}$ denote the indicator function.\\
Define $Z_{n,i}':=Z_{n,i}^*I{\{\sum_{j=1}^{i-1}{Z_{n,j}^*}^2\leq2V^*\}}$ and $S_{n,k}'=\sum_{i=1}^kZ_{n,i}',i=1,...,n$. Then, $\{S_{n,k}'\mid k=1,\dots,n\}$ is again a martingale with respect to $\{\mathcal{F}_{n,k}\mid k=1,\dots,n\}$ (for all $n$). Since
$$P^*(S_{n,n}'\neq T_{n,2}^{\text{mb}*})\leq P^*\bigg(\sum_{i=1}^n{Z_{n,i}^*}^2>2V^*\bigg)=o_P(1),$$
it is
\begin{equation}\label{boundedsumofsquares}
E^*[|\exp(\text{i}tS_{n,n}')-\exp(\text{i}tT_{n,2}^{\text{mb}*})|]=o_P(1).
\end{equation}
Thus, it is sufficient to prove $E^*[\exp(\text{i}tS_{n,n}')]=\exp\big(\frac{{V^*}^2t^2}{2}\big)+o_P(1)$.
For this purpose, define $r:\mathbb{R}\rightarrow\mathbb{C},x\mapsto \text{i}x+\frac{x^2}{2}-\log(1+\text{i}x)$,
\begin{eqnarray*}
R_n'(t)&=&\prod_{j=1}^n(1+\text{i}tZ_{n,j}')\\
W_n'(t)&=&\exp\bigg(-\frac{t^2}{2}\sum_{j=1}^nZ_{n,j}'^2+\sum_{j=1}^nr(tZ_{n,j}')\bigg)\\
J_n&=&\left\{
\begin{array}{ll}\min\Big(i\in\{1,...,n\}:\sum_{j=1}^i{Z^*_{n,j}}^2>2V^*\Big),&\textup{if }\sum_{j=1}^n{Z^*_{n,j}}^2>2V^*\\n,&\textup{otherwise},\end{array}\right.
\end{eqnarray*}
so that
\begin{eqnarray*}
\exp(\text{i}tS_{n,n}')&=&\exp\bigg(\text{i}t\sum_{j=1}^nZ_{n,j}'\bigg) \;=\;R_n'(t)W_n'(t)
\\&=&R_n'(t)\exp\bigg(-\frac{{V^*}^2t^2}{2}\bigg)+R_n'(t)\bigg(W_n'(t)-\exp\bigg(-\frac{{V^*}^2t^2}{2}\bigg)\bigg).
\end{eqnarray*}
Using Lemma \ref{lemmahallheyde} this leads to
\begin{eqnarray*}
E^*[|R_{n}'(t)|^2]&=&E^*\bigg[\prod_{i=1}^n(1+t^2Z_{n,i}'^2)\bigg]
\\&\leq&E^*\bigg[\exp\bigg(t^2\sum_{i=1}^{J_n-1}Z_{n,i}'^2\bigg)(1+t^2Z_{n,J_n}'^2)\bigg]
\\&\leq&\exp(2{V^*}t^2)(1+t^2E^*[Z_{n,J_n}'^2])\;\leq\;\exp(2{V^*}t^2)+o_P(1)
\end{eqnarray*}
applying (\ref{csu}). 
This in turn implies the uniform integrability of $R_n'(t)$ with respect to $P^*$, since
$$E^*\big[|R_n'(t)|I{\{|R_n'(t)|>C\}}\big]\leq\frac{E^*[|R_n'(t)|^2]}{C}\leq\frac{\exp(2{V^*}t^2)}{C}+o_P(1).$$
Moreover, as a consequence of the martingale property, one has $E[R_n'(t)]=1$. Due to
$$\bigg|R_n'(t)\bigg(W_n'(t)-\exp\bigg(-\frac{{V^*}^2t^2}{2}\bigg)\bigg)\bigg|\leq|\exp(itS_{n,n}')|+|R_n'(t)|\exp\bigg(-\frac{{V^*}^2t^2}{2}\bigg)$$
and $|\exp(itS_{n,n}')|=1$ the uniform integrability is transferred to the left hand side. For all $C_1,C_2>0$, one has
\begin{align*}
&\bigg|E^*[\exp(\text{i}tS_{n,n}')]-\exp\bigg(-\frac{{V^*}^2t^2}{2}\bigg)\bigg|
\\[0,2cm]&\quad\leq E^*\bigg[\bigg|R_n'(t)\bigg(W_n'(t)-\exp\bigg(-\frac{{V^*}^2t^2}{2}\bigg)\bigg)\bigg|\bigg]
\\[0,2cm]&\quad\leq E^*\bigg[|R_n'(t)|\bigg|W_n'(t)-\exp\bigg(-\frac{{V^*}^2t^2}{2}\bigg)\bigg|I{\{|R_n'(t)||W_n'(t)-\exp(-\frac{{V^*}^2t^2}{2})|>C_1\}}\bigg]
\\[0,2cm]&\quad\quad+E^*\bigg[|R_n'(t)|\bigg|W_n'(t)-\exp\bigg(-\frac{{V^*}^2t^2}{2}\bigg)\bigg|I{\{|R_n'(t)||W_n'(t)-\exp(-\frac{{V^*}^2t^2}{2})|\leq C_1\}}I{\{|R_n'(t)|\leq C_2\}}\bigg]
\\[0,2cm]&\quad\quad+\frac{C_1}{C_2}E^*\bigg[|R_n'(t)|I{\{|R_n'(t)|>C_2\}}\bigg].
\end{align*}
Now use the uniform integrability together with $W_n'(t)-\exp\big(-\frac{V^2t^2}{2}\big)=o_p(1)$ to obtain
$$E^*[\exp(\text{i}tS_{n,n}')]=\exp\bigg(-\frac{{V^*}^2t^2}{2}\bigg)+o_P(1).$$
\hfill $\Box$

\subsection{Proof of Theorem \ref{theo2}}\label{prooftheo2}

Using a Taylor expansion for $\Lambda_{\hat{\theta}}(Y)$ around $\Lambda_{\theta_0}(Y)$ one obtains the decomposition $I_n(\hat\theta)=I_n(\theta_0)+I+II+III+IV+V$, where
\begin{eqnarray*}
I &=& \frac{h^{\frac{p}{2}}}{n^3g^{2p}h^p}{\sum_{i,j,k,l}}^{\neq}
(\hat{\theta}-\theta_0)^ta_{i,k}^ta_{j,l}^t(\hat{\theta}-\theta_0)
L_{i,k}L_{j,l}K_{i,j}\psi_{i,j}\\
II &=& \frac{h^{\frac{p}{2}}}{n^3g^{2p}h^p}{\sum_{i,j,k,l}}^{\neq}
(\hat{\theta}-\theta_0)^tA_{i,k}^*(\hat{\theta}-\theta_0)(\hat{\theta}-\theta_0)^tA_{j,l}^*(\hat{\theta}-\theta_0)
L_{i,k}L_{j,l}K_{i,j}\psi_{i,j}\\
III&=& \frac{2h^{\frac{p}{2}}}{n^3g^{2p}h^p}{\sum_{i,j,k,l}}^{\neq}
(\Lambda_{\theta_0}(Y_i)-\Lambda_{\theta_0}(Y_k))a_{j,l}(\hat{\theta}-\theta_0)
L_{i,k}L_{j,l}K_{i,j}\psi_{i,j}\\
IV&=& 	\frac{2h^{\frac{p}{2}}}{n^3g^{2p}h^p}{\sum_{i,j,k,l}}^{\neq}
(\Lambda_{\theta_0}(Y_i)-\Lambda_{\theta_0}(Y_k))(\hat{\theta}-\theta_0)^tA_{j,l}^*(\hat{\theta}-\theta_0)
L_{i,k}L_{j,l}K_{i,j}\psi_{i,j}\\
V&=& 	\frac{2h^{\frac{p}{2}}}{n^3g^{2p}h^p}{\sum_{i,j,k,l}}^{\neq}
a_{i,k}(\hat{\theta}-\theta_0)(\hat{\theta}-\theta_0)^t A_{j,l}^* (\hat{\theta}-\theta_0)L_{i,k}L_{j,l}K_{i,j}\psi_{i,j}
\end{eqnarray*}
with $a_{i,k}=(\dot{\Lambda}_{\theta_0}(Y_i)-\dot{\Lambda}_{\theta_0}(Y_k))$ and $A_{i,k}^*=\big(\ddot\Lambda_{\theta_i^*}(Y_i)-\ddot\Lambda_{\theta_k^*}(Y_k)\big)$
for suitable $\theta_i^*$ between $\theta_0$ and $\hat{\theta}$. Further, we use the notations
	$$L_{i,k}=L\bigg(\frac{W_i-W_k}{g}\bigg),K_{i,j}=K\bigg(\frac{W_i-W_j}{h}\bigg),\psi_{i,j}=\psi(X_i-X_j).$$
Now for the first term one obtains
	\begin{align*}
	|I|&\leq\|\hat{\theta}-\theta_0\|^2\frac{h^{\frac{p}{2}}}{n^3g^{2p}h^p}{\sum_{i,j,k,l}}^{\neq}\|a_{i,k}\|\,\|a_{j,l}\|
	\\[0,2cm]&\quad\quad \Big|L\bigg(\frac{W_i-W_k}{g}\bigg)L\bigg(\frac{W_j-W_l}{g}\bigg)K\bigg(\frac{W_i-W_j}{h}\bigg)\psi(X_i-X_j)\Big|
	\end{align*}
with 
	\begin{align*}
	&\frac{1}{g^{2p}h^p}E\bigg[\|a_{1,3}\|\,\|a_{2,4}\|\Big|L\bigg(\frac{W_1-W_3}{g}\bigg)L\bigg(\frac{W_2-W_4}{g}\bigg)K\bigg(\frac{W_1-W_3}{h}\bigg)\psi(X_1-X_2)\Big|\bigg]
	\\[0,2cm]&\quad\leq\frac{C}{g^{2p}h^p}E\bigg[\Big|L\bigg(\frac{W_1-W_3}{g}\bigg)L\bigg(\frac{W_2-W_4}{g}\bigg)K\bigg(\frac{W_1-W_3}{h}\bigg)\Big|
	\\[0,2cm]&\quad\quad\quad E\big[\|\dot{\Lambda}_{\theta_0}(Y_1)-\dot{\Lambda}_{\theta_0}(Y_3)\|\,\|\dot{\Lambda}_{\theta_0}(Y_2)-\dot{\Lambda}_{\theta_0}(Y_4)\|\,\big|X_1,W_1,...,X_4,W_4\big]\bigg]
	\\[0,2cm]&\quad\leq\frac{C^2}{g^{2p}h^p}E\bigg[\Big|L\bigg(\frac{W_1-W_3}{g}\bigg)L\bigg(\frac{W_2-W_4}{g}\bigg)K\bigg(\frac{W_1-W_3}{h}\bigg)\Big|\bigg]
	\\[0,2cm]&\quad=O(1)	
\end{align*}
for some constant $C$, where for the first inequality we use assumption \ref{B7} and for the second inequality we use assumption \ref{B11}. The rate $O(1)$ follows by standard kernel arguments.  
Applying \ref{B10} we obtain $I=O_P(h^{\frac{p}{2}})=o_P(1)$.

Due to \ref{B10} we can assume that $\|\hat{\theta}-\theta\|\leq\alpha$ for $\alpha$ from \ref{B11}. Then $|II|$ can be bounded by 
	\begin{align*}
	&\frac{C\|\hat{\theta}-\theta_0\|^4}{n^3h^{\frac{p}{2}}g^{2p}}{\sum_{i,k,j,l}}^{\neq}\|A_{i,k}^*\|\|A_{j,l}^*\| |L_{i,k}L_{j,l}|
	\\[0,2cm]&\leq\frac{Cn\|\hat{\theta}-\theta_0\|^4}{h^{\frac{p}{2}}g^{2p}}\bigg(\frac{1}{n^2}\sum_{i\neq k}\|A_{i,k}^*\|L_{i,k}\bigg)^2	
	\\[0,2cm]&\leq\frac{4Cn\|\hat{\theta}-\theta_0\|^4}{h^{\frac{p}{2}}}\bigg(\frac{1}{n}\sum_{i=1}^n\underset{\|\theta-\theta_0\|\leq\alpha}{\sup}\,\|\ddot\Lambda_{\theta}(Y_i)\|\sup_{w}\frac{1}{n}\sum_{k=1}^n\frac{1}{g^p}\Big|L\bigg(\frac{W_i-w}{g}\bigg)\Big|\bigg)^2
	\\[0,2cm]&=O_P\bigg(\frac{1}{nh^{\frac{p}{2}}}\bigg) \;=\; o_P(1). 
		\end{align*}
		Here, the uniform convergence of a kernel density estimator with kernel $|L|$ and bandwidth $g$ is used.

The treatments of $IV$ and $V$ are similar and omitted for the sake of brevity. 

For $III$ we need a further decomposition that is obtained by using the model equation (\ref{model}) and the structure of the local alternative. We have  $III=A_n-B_n+C_n+D_n$ with
\begin{eqnarray*}
A_n &=& \frac{2h^{\frac{p}{2}}}{n^3g^{2p}h^p}{\sum_{i,j,k,l}}^{\neq}
\eps_ia_{j,l}L_{i,k}L_{j,l}K_{i,j}\psi_{i,j}(\hat{\theta}-\theta_0)\\
B_n &=& \frac{2h^{\frac{p}{2}}}{n^3g^{2p}h^p}{\sum_{i,j,k,l}}^{\neq}
\eps_k a_{j,l}L_{i,k}L_{j,l}K_{i,j}\psi_{i,j}(\hat{\theta}-\theta_0)\\
C_n&=& \frac{2h^{\frac{p}{2}}\delta_n}{n^3g^{2p}h^p}{\sum_{i,j,k,l}}^{\neq}
(d(W_i,V_i)-d(W_k,V_k))a_{j,l}
L_{i,k}L_{j,l}K_{i,j}\psi_{i,j}(\hat{\theta}-\theta_0)\\
D_n &=& \frac{2h^{\frac{p}{2}}}{n^3g^{2p}h^p}{\sum_{i,j,k,l}}^{\neq}
(r(W_i)-r(W_k))a_{j,l}
L_{i,k}L_{j,l}K_{i,j}\psi_{i,j}(\hat{\theta}-\theta_0).
\end{eqnarray*}
To treat $A_n$ define the function $\varphi(x,y)=g^{-2p}h^{-p}E[a_{1,2}L_{3,4}L_{1,2 }K_{3,1}\psi_{3,1 }|X_3=x,Y_3=y]$ and consider $A_n=A_n^{(1)}+A_n^{(2)}$ with 
\begin{eqnarray*}
A_n^{(1)} &=& O(h^{\frac{p}{2}})\sum_{i=1}^n \eps_i \varphi(X_i,Y_i)(\hat{\theta}-\theta_0) \;=\; O(h^{\frac{p}{2}})O_P(n^{1/2})O_P(n^{-1/2})\;=\; o_P(1)
\end{eqnarray*}
because the random variables $\eps_i \varphi(X_i,Y_i)$, $i=1,\dots,n$, are iid and centered, while the function $\varphi$ is bounded.
Further, one obtains 
\begin{eqnarray*}
A_n^{(2)} &=& O(h^{\frac{p}{2}})\frac{1}{n^3}{\sum_{i,j,k,l}}^{\neq}
\eps_i \Big(a_{j,l}L_{i,k}L_{j,l}K_{i,j}\psi_{i,j}-E[a_{j,l}L_{i,k}L_{j,l}K_{i,j}\psi_{i,j}|X_i,Y_i]\Big)(\hat{\theta}-\theta_0) \;=\; o_P(1)
\end{eqnarray*}
with some tedious calculation of the second moment of the vector components of the (centered) sum (without the factor $(\hat{\theta}-\theta_0)$). 

For $C_n$ note that $E[g^{-2p} h^{-p}|(d(W_i,V_i)-d(W_k,V_k))a_{j,l}
L_{i,k}L_{j,l}K_{i,j}\psi_{i,j}|]<\infty$ and thus
$$C_n= O(h^{p/2}\delta_n n )O_P(1)O_P(n^{-1/2})=O_P(h^{p/4})=o_P(1).$$

It remains to show $D_n=o_P(1)$, which can be done by calculating expectation and variance of the sum (without the factor $(\hat{\theta}-\theta_0)$). As, in contrast to $B_n$, the summands are not centered, to this end one needs to make use of assumptions \ref{B1} and \ref{B4}. Consider, e.g., for $i\neq k$
\begin{eqnarray*}
E[(r(W_i)-r(W_k))g^{-p}L_{i,k}|W_i=w] &=& \int (r(w)-r(v))f_W(v)\frac{1}{g^p}L\bigg(\frac{w-v}{g}\bigg)\,dv\\
&=& \int (r(w)-r(w-ug))f_W(w-ug) L(u)\,du \\
&=& O(g^2) +O(g)I\{w\in B_g\}\;=\; O(g^2)
\end{eqnarray*}
using Taylor's expansions for $f_W$ and $rf_W$ and the order of the kernel $L$. Here $B_g$ denotes the $g$-boundary of the support of $W$ and one has $P(W\in B_g)=O(g)$. 
Thus, the expectation of the sum (without the factor $(\hat{\theta}-\theta_0)$) is of order $h^{p/2}ng^2$ with $h^{p/2}ng^2n^{-1/2}=o(1)$ by assumption \ref{B8}. The calculation of the variance is more tedious, but analogous to the calculations in Lavergne et al.\ (2015) and Lavergne and Vuong (2000).
\hfill $\Box$

\section*{References}

\begin{description}

\item Alcalß, J.T., Crist¾bal, J.A. and Gonzßlez-Manteiga, W. (1999). Goodness-of-fit test for linear models
based on local polynomials. {\it Statist. Prob. Lett.} {\bf  42}, 39--46. 

\item Allison, J.S., Hu\v{s}kovß, M. and Meintanis, S.G. (2018). Testing the adequacy of semiparametric transformation models. {\it TEST} {\bf 27}, 70--94. 

\item Berk, R., Brown, L., Buja, A., Zhang, K. and Zhao, L. (2013). 
Valid post-selection inference. {\it Ann. Statist.} {\bf 41}, 802--837. 

\item Bierens, H.J. (1982). Consistent model specification tests. {\it J. Econometr.} {\bf 20}, 105--134. 

\item Bickel, P. and Doksum, K.A. (1981). An analysis of transformations revisited.
{\it J. Amer. Statist. Assoc.} {\bf  76}, 296--311. 

\item Box, G.E.P. and Cox, D.R. (1964).
An analysis of transformations.
{\it J. Roy. Statist. Soc. Ser. B}  {\bf 26}, 211--252.

\item  B³cher, A. and Dette, H. (2013).  Multiplier bootstrap of tail copulas with applications. {\it Bernoulli} {\bf  19}, 1655--1687.

\item  Carroll, R. J. and Ruppert, D. (1988).
\textit{Transformation and Weighting in Regression}. 
Monographs on Statistics and Applied Probability. 
Chapman \& Hall, New York.

\item Charkhi, A. and Claeskens, G. (2018). Asymptotic post-selection inference for the Akaike information criterion. {\it 
Biometrika} {\bf  105}, 645--664. 

\item Colling, B. and Van Keilegom, I. (2016).
Goodness-of-fit tests in semiparametric transformation models. 
 {\it TEST} {\bf 25}, 291--308. 

\item Colling, B. and Van Keilegom, I. (2017).
Goodness-of-fit tests in semiparametric transformation models using the integrated regression function. 
 {\it J. Multivariate Anal.} {\bf 160}, 10--30. 
 
 \item Colling, B. and Van Keilegom, I. (2018).
Estimation of a semiparametric transformation model: a novel approach based on least squares minimization. preprint KU Leuven available at https://limo.libis.be/primo-explore/search?vid=Lirias

\item Delgado, M. and Gonzßlez-Manteiga, W. (2001). Significance testing in nonparametric regression based on the bootstrap. 
{\it Ann. Statist.} {\bf 29}, 1469--1507. 

\item  Efron, B. (2014).  Estimation and accuracy after model selection. {\it J. Amer. Statist. Assoc.} {\bf 109}, 991--1007. 
 
\item Fan, J. and Gijbels, I (1996). \textit{Local Polynomial Modelling and its Applications}. Monographs on Statistics and Applied Probability. Chapman \& Hall, New York.

\item Fan, Y. and Li, Q (1996).  Consistent model specification tests: omitted variables and semiparametric functional
forms. {\it Econometrica} {\bf  64}, 865--890. 
 
\item Escanciano, J.C. (2006). A consistent test for regression models using projections.
{\it Econometric Theory} {\bf 22}, 1030--1051. 


\item Gonzßlez-Manteiga, W. and Crujeiras, R.M. (2013). An updated review of Goodness-of-Fit tests
for regression models. With discussion. {\it TEST} {\bf 22}, 361--411. 

\item Hall, P. and Heyde, C.C. (1980). {\it Martingale Limit Theory and Its Application.} Academic Press, New York. 

\item H\"ardle, W. and Mammen, E. (1993). Comparing Nonparametric Versus Parametric Regression Fits. \textit{Ann. Statist.} \textbf{21}, 1926--1947.

\item Hinkley, D.V. and Runger, G. (1984). The analysis of transformed data.
With discussion. {\it J. Amer. Statist. Assoc.} {\bf 79}, 302--320. 

\item Hlßvka, Z.,  Hu\v{s}kovß, M., Kirch, C. and  Meintanis, S.G. (2017). Fourier-type tests involving martingale difference processes.
{\it  Econometric Reviews} {\bf 36}, 468--492. 

\item Horowitz, J.L. (1996). Semiparametric estimation of a regression model with an unknown transformation
of the dependent variable. {\it Econometrica} {\bf 64}, 102--137. 

\item Horowitz, J.L. (2009).
\textit{Semiparametric and nonparametric methods in econometrics}.
Springer Series in Statistics. Springer, New York.

\item Hu\v{s}kovß, M. and Meintanis, S.G. (2009). 
Goodness-of-fit tests for parametric regression models based on empirical characteristic functions. {\it Kybernetika} {\bf 45}, 960--971. 

\item  Lavergne, P., Maistre, S. and Patilea, V. (2015). A significance test for covariates in nonparametric regression.
    {\it Electron. J. Statist.} {\bf 9}, 643--678. 

\item Lavergne, P. and Vuong, Q. (2000). Nonparametric significance testing. {\it Econometric Theory} {\bf 16}, 576--601.

\item Lee, J.D., Sun D.L., Sun, Y. and  Taylor, J.E. (2016). Exact post-selection inference, with application to the lasso. {\it Ann. Statist.} {\bf 44}, 907--927. 

\item Linton, O., Chen, R., Wang, N. and Hõrdle, W. (1997). An analysis of transformations for
additive nonparametric regression. {\it J. Amer. Statist. Assoc.} {\bf 92}, 1512--1521. 

\item Linton, O., Sperlich, S. and Van Keilegom, I. (2008).
Estimation on a semiparametric transformation model.
{\it Ann. Statist.}  {\bf 36}, 686--718.

\item Nadaraya, E. A. (1964). On Estimating Regression. {\it Theory of Probability and Its Applications} {\bf 9}, 141--142.

\item R Core Team (2013). R: A language and environment for statistical
  computing. R Foundation for Statistical Computing, Vienna, Austria.
  URL http://www.R-project.org/
  
\item Stute, W. (1997).  Nonparametric model checks for regression. {\it Ann. Statist.} {\bf 25}, 613--641. 

\item van der Vaart, A.W. and Wellner, J A. (1996).
\textit{Weak Convergence and Empirical Processes}.
Springer-Verlag, New York.

\item Van Keilegom, I., Gonzßlez-Manteiga, W. and Sßnchez Sellero, C. (2008). Goodness of fit tests in parametric regression based on the estimation of the error distribution. {\it TEST} {\bf 17}, 401--415.

\item Watson, G.S. (1964). Smooth Regression Analysis. {\it The Indian Journal of Statistics} {\bf 26}, 359--372.

\item Wu, C.F.J. (1986). Jackknife, bootstrap and other resampling methods in regression analysis. {\it Ann. Statist.} {\bf 14}, 1261--1350.

\item Yeo, I-K. and Johnson, R. A. (2000).
A new family of power transformations to improve normality or symmetry. 
{\it Biometrika}  {\bf 87}, 954--959.

\item Zheng, J.X. (1996). A consistent test of functional form via nonparametric
estimation techniques. {\it Journal of Econometrics} {\bf  75}, 263--289. 

\item Zhu, L., Fujikoshi, N. and Naito, K. (2001). Heteroscedasticity checks for regression models. {\it Science in China Series A} {\bf 44}, 1237--1252.

\item Zellner, A. and Revankar, N.S. (1969). Generalized production functions. {\it Rev. Economic
Studies} {\bf 36}, 241--250. 

\end{description}
\end{appendix}
\newpage
\pagestyle{empty}

\begin{table}[p]\footnotesize
	\centering
	\begin{tabular}{|l|r|r|r|r|r|r|r|r|r|r|r|r|}
		\hline
		Alternative&\multicolumn{5}{c|}{$T_n(\hat\theta)$}&\multicolumn{5}{c|}{$V_n(\hat\theta)$}&$W_{\exp}^2$&max\\
		\hline
		&swb&twb&mb&cmb&asym&swb&twb&mb&cmb&asym&twb&twb\\
			\hline
			$0$&0.162&0.108&0.038&0.088&0.048&0.156&0.116&0.044&0.096&0.050&0.122&0.134\\
			\hline
			$2X^2$&0.428&0.312&0.320&0.452&0.328&0.524&0.416&0.338&0.444&0.344&0.500&0.512\\
			$3X^2$&0.698&0.578&0.604&0.730&0.620&0.776&0.682&0.624&0.726&0.628&0.788&0.790\\
			$4X^2$&0.882&0.786&0.842&0.902&0.846&0.922&0.868&0.852&0.902&0.856&0.916&0.916\\
			$5X^2$&0.962&0.914&0.952&0.976&0.958&0.982&0.944&0.954&0.972&0.954&0.944&0.946\\
			\hline
			$2\exp(X)$&0.378&0.240&0.256&0.388&0.266&0.472&0.314&0.282&0.390&0.294&0.382&0.392\\
			$3\exp(X)$&0.598&0.360&0.508&0.624&0.510&0.674&0.492&0.526&0.626&0.532&0.608&0.608\\
			$4\exp(X)$&0.776&0.514&0.704&0.804&0.712&0.852&0.638&0.716&0.792&0.726&0.716&0.732\\
			$5\exp(X)$&0.888&0.630&0.852&0.900&0.858&0.914&0.750&0.854&0.896&0.856&0.810&0.812\\
			\hline
			$0.25\sin(2\pi X)$&0.326&0.234&0.146&0.238&0.146&0.338&0.270&0.158&0.228&0.160&0.124&0.294\\
			$0.5\sin(2\pi X)$&0.718&0.644&0.508&0.656&0.526&0.774&0.722&0.518&0.646&0.532&0.134&0.590\\
			$0.75\sin(2\pi X)$&0.982&0.948&0.924&0.968&0.932&0.986&0.976&0.916&0.960&0.922&0.160&0.924\\
			$1\sin(2\pi X)$&1.000&1.000&0.996&1.000&0.998&1.000&1.000&0.994&0.998&0.994&0.178&0.998\\
			\hline
	\end{tabular}
	\caption{Rejection probabilities at $\theta_0=0$ for the lack-of-fit tests $T_n(\hat\theta)$, $V_n(\hat\theta)$ with standard wild bootstrap (swb), transformation wild bootstrap (twb), multiplier bootstrap (mb), centered multiplier bootstrap (cmb), using asymptotic critical values (asym),  and for $W_{\exp}^2$ (twb) with nominal level $0.1$. Moreover, the maximum rejection probability of the tests corresponding to $T_{CM},W_{1}^2,W_{\exp_i}^2,W_{\exp}^2,W_{1/\exp}^2,W_{\sin}^2$ (twb) is displayed.}
	\label{table1}
\end{table}

\begin{table}[htbp]\footnotesize
	\centering
		\begin{tabular}{|l|r|r|r|r|r|r|r|r|r|r|r|r|}
			\hline
			Alternative&\multicolumn{5}{c|}{$T_n(\hat\theta)$}&\multicolumn{5}{c|}{$V_n(\hat\theta)$}&$W_{\exp}^2$&max\\
			\hline
		&swb&twb&mb&cmb&asym&swb&twb&mb&cmb&asym&twb&twb\\
			\hline
			$0$&0.182&0.082&0.058&0.118&0.068&0.190&0.134&0.066&0.124&0.064&0.128&0.128\\
			\hline
			$2X^2$&0.440&0.248&0.296&0.438&0.322&0.514&0.400&0.320&0.436&0.334&0.504&0.504\\
			$3X^2$&0.690&0.512&0.588&0.718&0.602&0.764&0.656&0.614&0.712&0.622&0.788&0.788\\
			$4X^2$&0.872&0.702&0.834&0.888&0.836&0.908&0.844&0.834&0.888&0.846&0.914&0.914\\
			$5X^2$&0.964&0.882&0.954&0.974&0.960&0.982&0.946&0.960&0.972&0.960&0.970&0.970\\
			\hline
			$2\exp(X)$&0.386&0.156&0.262&0.386&0.274&0.460&0.310&0.286&0.390&0.286&0.322&0.330\\
			$3\exp(X)$&0.608&0.230&0.508&0.618&0.510&0.678&0.484&0.520&0.612&0.526&0.528&0.536\\
			$4\exp(X)$&0.772&0.282&0.682&0.784&0.692&0.840&0.616&0.704&0.776&0.708&0.648&0.676\\
			$5\exp(X)$&0.874&0.304&0.838&0.884&0.844&0.908&0.718&0.850&0.882&0.854&0.726&0.744\\
			\hline
			$0.25\sin(2\pi X)$&0.322&0.162&0.138&0.220&0.148&0.324&0.260&0.154&0.222&0.158&0.120&0.288\\
			$0.5\sin(2\pi X)$&0.762&0.506&0.544&0.680&0.566&0.790&0.744&0.556&0.678&0.566&0.128&0.566\\
			$0.75\sin(2\pi X)$&0.962&0.856&0.922&0.954&0.922&0.976&0.972&0.924&0.954&0.928&0.128&0.844\\
			$1\sin(2\pi X)$&1.000&0.976&0.996&0.998&0.996&1.000&1.000&0.996&0.996&0.996&0.138&0.980\\
			\hline
		\end{tabular}
	\caption{Rejection probabilities at $\theta_0=0.5$ for the lack-of-fit tests $T_n(\hat\theta)$, $V_n(\hat\theta)$ with standard wild bootstrap (swb), transformation wild bootstrap (twb), multiplier bootstrap (mb), centered multiplier bootstrap (cmb), using asymptotic critical values (asym)  and for $W_{\exp}^2$ (twb) with nominal level $0.1$. Moreover, the maximum rejection probability of the tests corresponding to $T_{CM},W_{1}^2,W_{\exp_i}^2,W_{\exp}^2,W_{1/\exp}^2,W_{\sin}^2$ (twb) is displayed.}
	\label{table2}
\end{table}

\begin{table}[htbp]\footnotesize
	\centering
		\begin{tabular}{|l|r|r|r|r|r|r|r|r|r|r|r|r|}
			\hline
			Alternative&\multicolumn{5}{c|}{$T_n(\hat\theta)$}&\multicolumn{5}{c|}{$V_n(\hat\theta)$}&$W_{\exp}^2$&max\\
			\hline
		&swb&twb&mb&cmb&asym&swb&twb&mb&cmb&asym&twb&twb\\
			\hline
			$0$&0.166&0.104&0.050&0.108&0.056&0.182&0.124&0.058&0.110&0.056&0.106&0.108\\
			\hline
			$2X^2$&0.426&0.284&0.286&0.432&0.304&0.504&0.394&0.304&0.430&0.322&0.492&0.492\\
			$3X^2$&0.694&0.552&0.584&0.726&0.602&0.770&0.662&0.618&0.718&0.630&0.774&0.788\\
			$4X^2$&0.878&0.746&0.832&0.894&0.840&0.920&0.848&0.838&0.894&0.844&0.904&0.904\\
			$5X^2$&0.962&0.890&0.948&0.972&0.954&0.982&0.940&0.954&0.970&0.954&0.964&0.964\\
			\hline
			$2\exp(X)$&0.376&0.224&0.242&0.368&0.250&0.442&0.300&0.268&0.370&0.268&0.294&0.300\\
			$3\exp(X)$&0.594&0.322&0.496&0.608&0.498&0.674&0.486&0.510&0.604&0.514&0.470&0.476\\
			$4\exp(X)$&0.774&0.424&0.688&0.782&0.698&0.848&0.626&0.706&0.782&0.710&0.618&0.624\\
			$5\exp(X)$&0.884&0.486&0.848&0.894&0.852&0.920&0.736&0.856&0.892&0.858&0.676&0.692\\
			\hline
			$0.25\sin(2\pi X)$&0.320&0.198&0.136&0.222&0.144&0.330&0.264&0.152&0.226&0.154&0.104&0.278\\
			$0.5\sin(2\pi X)$&0.752&0.546&0.542&0.678&0.560&0.790&0.746&0.556&0.678&0.568&0.106&0.554\\
			$0.75\sin(2\pi X)$&0.962&0.882&0.918&0.954&0.920&0.976&0.972&0.924&0.954&0.926&0.112&0.840\\
			$1\sin(2\pi X)$&1.000&0.968&0.996&0.998&0.996&1.000&1.000&0.996&0.996&0.996&0.116&0.982\\
			\hline
		\end{tabular}
	\caption{Rejection probabilities at $\theta_0=1$ for the lack-of-fit tests $T_n(\hat\theta)$, $V_n(\hat\theta)$ with standard wild bootstrap (swb), transformation wild bootstrap (twb), multiplier bootstrap (mb), centered multiplier bootstrap (cmb), using asymptotic critical values (asym)  and for $W_{\exp}^2$ (twb) with nominal level $0.1$. Moreover, the maximum rejection probability of the tests corresponding to $T_{CM},W_{1}^2,W_{\exp_i}^2,W_{\exp}^2,W_{1/\exp}^2,W_{\sin}^2$ (twb) is displayed.}
	\label{table3}
\end{table}

\begin{table}[p]\footnotesize
	\centering
	\begin{small}
		\begin{tabular}{|l|r|r|r|r|r|r|r|r|r|r|r|r|}
			\hline
			&\multicolumn{6}{c|}{$n=75$}&\multicolumn{6}{c|}{$n=100$}\\
			\hline
			Model&mb&cmb&swb&twb&$\Psi_{n,1}$&$\tilde{\Psi}_n$&mb&cmb&swb&twb&$\Psi_{n,1}$&$\tilde{\Psi}_n$\\
			\hline
			(\ref{null})&0.016&0.028&0.048&0.036&0.030&0.050&0.012&0.034&0.058&0.038&0.030&0.040\\
			(\ref{V})&0.216&0.348&0.366&0.230&0.470&0.490&0.324&0.436&0.474&0.324&0.600&0.600\\
			(\ref{0.25sin(2piV)})&0.084&0.168&0.170&0.124&0.100&0.100&0.128&0.214&0.238&0.120&0.110&0.110\\
			(\ref{sin(5V)})&0.904&0.918&0.968&0.868&0.790&0.770&0.978&0.980&0.994&0.948&0.809&0.880\\
			(\ref{5V2})&0.998&0.998&1.000&0.744&0.390&0.410&1.000&1.000&0.998&0.850&0.460&0.510\\
			(\ref{V2})&0.240&0.360&0.384&0.196&0.300&0.340&0.356&0.470&0.522&0.304&0.380&0.400\\
			(\ref{exp(V2)})&0.598&0.696&0.724&0.410&0.300&0.350&0.762&0.824&0.864&0.507&0.400&0.440\\
			\hline
		\end{tabular}
	\end{small}
	\caption{Rejection probabilities for the test of significance with test statistic $\tilde{I}_n(\hat\theta)$  with  multiplier bootstrap (mb) and centered multiplier bootstrap (cmb), test statistic  $I_n(\hat\theta)$ with standard wild bootstrap (swb) and transformation wild bootstrap (twb) and  test statistics $\Psi_{n,1},\tilde{\Psi}_n$ from Allison et al.\ with transformation wild bootstrap. The transformation parameter is $\theta_0=1$ and the nominal level is $0.05$.}
	\label{table5}
\end{table}

	\begin{table}[htbp]\footnotesize
		\centering
		\begin{small}
			\begin{tabular}{|l|r|r|r|r|r|r|r|r|r|}
				\hline
				&\multicolumn{8}{c|}{$\alpha=0.05$}&$\bar{h}$\\
				\hline
				&\multicolumn{2}{c|}{$h=cv$}&\multicolumn{2}{c|}{$h=2cv$}&\multicolumn{2}{c|}{$h=0.5cv$}&\multicolumn{2}{c|}{$h=0.2$}&$h=cv$\\
				\hline
				Model&mb&cmb&mb&cmb&mb&cmb&mb&cmb&\\
				\hline
				(\ref{null})&0.012&0.034&0.014&0.048&0.004&0.016&0.012&0.018&0.4039648\\
				(\ref{V})&0.362&0.494&0.436&0.568&0.266&0.382&0.298&0.416&0.4206143\\
				(\ref{0.25sin(2piV)})&0.168&0.250&0.194&0.278&0.100&0.172&0.112&0.202&0.4343091\\
				(\ref{sin(5V)})&0.962&0.974&0.988&0.990&0.926&0.942&0.994&0.998&0.420002\\
				(\ref{5V2})&1.000&1.000&1.000&1.000&0.994&0.996&1.000&1.000&0.3979705\\
				(\ref{V2})&0.386&0.506&0.468&0.576&0.284&0.406&0.320&0.454&0.4208178\\
				(\ref{exp(V2)})&0.768&0.816&0.838&0.892&0.654&0.730&0.762&0.838&0.4095975\\
				\hline
			\end{tabular}
		\end{small}
		\caption{Rejection probabilities for the test of significance with test statistic $\tilde{I}_n(\hat\theta)$ with mb and cmb for the bandwidths $h=cv,h=2cv,h=0.5cv$ and $h=0.2$, where $cv$ is the bandwidth obtained via cross validation. The transformation parameter is equal to $\theta_0=1$, the sample size is equal to $n=100$ and the level is equal to $\alpha=0.05$.}
		\label{table6}
	\end{table}

	\begin{table}[htbp]\footnotesize
	\centering
	\begin{small}
		\begin{tabular}{|l|r|r|r|r|r|r|}
			\hline
			&\multicolumn{2}{c|}{$v=0.05$}&\multicolumn{2}{c|}{$v=0.1$}&\multicolumn{2}{c|}{$v=0.2$}\\
			\hline
			Model&mb&cmb&mb&cmb&mb&cmb\\
			\hline
			(\ref{null})&0.012&0.028&0.012&0.034&0.006&0.036\\
			(\ref{V})&0.306&0.454&0.362&0.494&0.342&0.460\\
			(\ref{0.25sin(2piV)})&0.146&0.238&0.168&0.250&0.118&0.200\\
			(\ref{sin(5V)})&0.960&0.970&0.962&0.974&0.954&0.964\\
			(\ref{5V2})&1.000&1.000&1.000&1.000&1.000&1.000\\
			(\ref{V2})&0.328&0.482&0.386&0.506&0.342&0.478\\
			(\ref{exp(V2)})&0.732&0.814&0.768&0.816&0.740&0.794\\
			\hline
		\end{tabular}
	\end{small}
	\caption{Rejection probabilities and cross validation bandwidths for the test of significance with test statistic $\tilde{I}_n(\hat\theta)$ with mb and cmb for different $\psi$ ($v=0.05,v=0.1$ and $v=0.2$). The transformation parameter is equal to $\theta_0=1$, the sample size is equal to $n=100$ and the level is equal to $\alpha=0.05$.}
	\label{table7}
\end{table}

\end{document}